\def\isabridged{1}
\def\isarxiv{1}

\newif\ifarxiv
\newif\ifnotarxiv

\ifdefined\isarxiv
\arxivtrue
\fi

\ifarxiv\notarxivfalse
\else\notarxivtrue
\fi

\ifarxiv
\PassOptionsToPackage{dvipsnames}{xcolor}
\fi
\documentclass[sigconf]{acmart}
\usepackage[dvipsnames]{xcolor}
\usepackage{microtype}
\usepackage{url}
\usepackage{listings}
\usepackage{todonotes}
\usepackage{glossaries}
\usepackage{bytefield}
\usepackage{pifont}
\usepackage{xspace}
\usepackage{cleveref}
\usepackage{paralist}
\usepackage{booktabs}
\usepackage{censor}
\usepackage{enumitem}
\usepackage{colortbl}
\usepackage{tikz}
\usepackage{adjustbox}
\usepackage{subcaption}
\usepackage{multirow}
\usepackage[compact]{titlesec}

\glsdisablehyper
\newif\ifabridged
\newif\ifnotabridged
\newif\ifanonymous
\newif\ifnotanonymous

\ifdefined\isabridged
\abridgedtrue
\fi

\ifabridged\notabridgedfalse
\else\notabridgedtrue
\fi

\ifdefined\isanonymous
\anonymoustrue
\else\notanonymoustrue
\fi

\newcommand{\dOne}{\ding{182}\xspace}
\newcommand{\dTwo}{\ding{183}\xspace}
\newcommand{\dThree}{\ding{184}\xspace}
\newcommand{\dFour}{\ding{185}\xspace}
\newcommand{\dFive}{\ding{186}\xspace}

\newcommand{\dCOne}{\ding{192}\xspace}
\newcommand{\dCTwo}{\ding{193}\xspace}
\newcommand{\dCThree}{\ding{194}\xspace}
\newcommand{\dCFour}{\ding{195}\xspace}
\newcommand{\dCFive}{\ding{196}\xspace}
\newcommand{\dCSix}{\ding{197}\xspace}
\newcommand{\dCSeven}{\ding{198}\xspace}
\newcommand{\dCEight}{\ding{199}\xspace}

\newcommand{\xmark}{\ding{55}}

\Crefname{section}{\S$\!$}{\S\S$\!$}

\newsavebox{\bfbox}

\newcommand{\colorbitbox}[3]{
  \sbox0{\bitbox{#2}{#3}}
  \makebox[0pt][l]{\textcolor{#1}{\rule[-\dp0]{\wd0}{\ht0}}}
  \bitbox{#2}{#3}
}
\definecolor{mGreen}{rgb}{0,0.6,0}
\definecolor{mGray}{rgb}{0.5,0.5,0.5}
\definecolor{lGray}{rgb}{0.9,0.9,0.9}
\definecolor{mPurple}{rgb}{0.58,0,0.82}
\definecolor{backgroundColour}{rgb}{0.95,0.95,0.92}

\newcommand\LSTSize{\fontsize{7}{7.2}\selectfont}
\newcommand*\LSTfont{\LSTSize\ttfamily\SetTracking{encoding=*}{-0}\lsstyle}

\newcommand\realnumberstyle[1]{#1}

\makeatletter
\newcommand{\zebra}[2]{%
    {\realnumberstyle{#2}}%
    \begingroup
    \lst@basicstyle
    \ifodd\value{lstnumber}%
        \color{#1}%
        \rlap{\hspace*{\lst@numbersep}%
        \color@block{\linewidth}{\ht\strutbox}{\dp\strutbox}%
        }%
    \fi
    \endgroup
}
\makeatother

\makeatletter
\def\bstctlcite{\@ifnextchar[{\@bstctlcite}{\@bstctlcite[@auxout]}}
\def\@bstctlcite[#1]#2{\@bsphack
  \@for\@citeb:=#2\do{%
    \edef\@citeb{\expandafter\@firstofone\@citeb}%
    \if@filesw\immediate\write\csname #1\endcsname{\string\citation{\@citeb}}\fi}%
  \@esphack}
\makeatother

\lstdefinestyle{CStyle}{
    backgroundcolor=\color{backgroundColour},
    commentstyle=\color{mGreen},
    keywordstyle=\color{magenta},
    numberstyle=\tiny\zebra{lGray},
    stringstyle=\color{mPurple},
    basicstyle=\LSTfont,
    breakatwhitespace=false,
    breaklines=true,
    captionpos=b,    
    escapeinside={\%*}{*)},
    keepspaces=true,
    numbers=left,
    numbersep=5pt,
    showspaces=false,
    showstringspaces=false,
    showtabs=false,
    tabsize=2,
    language=C,
    classoffset=1,
    morekeywords={__blinded},
    keywordstyle=\color{orange},
    classoffset=2,
    classoffset=0,
}
\lstdefinelanguage[RISC-V]{Assembler}
{
  alsoletter={.},
  alsodigit={0x},
  morekeywords=[1]{
    lb, lh, lw, lbu, lhu,
    sb, sh, sw,
    sll, slli, srl, srli, sra, srai,
    add, addi, sub, lui, auipc,
    xor, xori, or, ori, and, andi,
    slt, slti, sltu, sltiu,
    beq, bne, blt, bge, bltu, bgeu,
    j, jr, jal, jalr, ret,
    scall, break, nop
  },
  morekeywords=[2]{
    .align, .ascii, .asciiz, .byte, .data, .double, .extern,
    .float, .globl, .half, .kdata, .ktext, .set, .space, .text, .word
  },
  morekeywords=[3]{
    zero, ra, sp, gp, tp, s0, fp,
    t0, t1, t2, t3, t4, t5, t6,
    s1, s2, s3, s4, s5, s6, s7, s8, s9, s10, s11,
    a0, a1, a2, a3, a4, a5, a6, a7,
    ft0, ft1, ft2, ft3, ft4, ft5, ft6, ft7,
    fs0, fs1, fs2, fs3, fs4, fs5, fs6, fs7, fs8, fs9, fs10, fs11,
    fa0, fa1, fa2, fa3, fa4, fa5, fa6, fa7
  },
  morecomment=[l]{;},
  morecomment=[l]{\#},
  morestring=[b]",
  morestring=[b]'
}
\definecolor{mauve}{rgb}{0.58,0,0.82}

\usetikzlibrary{decorations.pathreplacing,positioning,arrows,tikzmark}

\tikzset{
  comment/.style={
    draw=none,
    text=black,
    align=left,
    inner sep=2,
    outer sep=1,
    font=\rmfamily
  },
  none/.style={
    draw=none,
    text=black,
    align=left,
    inner sep=2,
    outer sep=1,
    font=\rmfamily
  },
}

\glsaddkey
    {hyphenated}
    {\relax}
    {\glsentryhyph}
    {\Glsentryhyph}
    {\glshyph}
    {\Glshyph}
    {\GLShyph}

\renewcommand{\paragraph}{\noindent\textbf}
\newacronym{abi}{ABI}{application binary interface}
\newacronym{ami}{AMi}{Architectural-Mimicry}
\newacronym{alu}{ALU}{arithmetic logic unit}
\newacronym{api}{API}{application programming interface}
\newacronym{blackout}{\texttt{BLACKOUT}}{Blinded Architectural Capabilities and Kernel for Oblivious Userspace Tasks}
\newacronym{brr}{BRR}{blinded register record}
\newacronym[longplural={\texttt{blinded} attributes}]{ba}{BA}{\texttt{blinded} attribute}
\newacronym[longplural={blinded capabilities},hyphenated={blinded-capability}]{bc}{BC}{blinded capability}
\newacronym[longplural={blinded memory}]{bd}{BD}{blinded memory}
\newacronym[longplural={blinded variables}]{bv}{BV}{blinded variable}
\newacronym[longplural={blinded registers}]{br}{BR}{blinded register}
\newacronym{btb}{BTB}{branch target buffer}
\newacronym{cisa}{CISA}{Cybersecurity & Infrastructure Security Agency}
\newacronym{cheri}{CHERI}{Capability Hardware Enhanced RISC Instructions}
\newacronym{clc}{\texttt{clc}}{load capability via capability}
\newacronym{cpu}{CPU}{central processing unit}
\newacronym{csc}{\texttt{csc}}{store capability via capability}
\newacronym{capSpecCon}{CSC}{Capability Speculation Contract}
\newacronym{csp}{\texttt{csp}}{stack pointer capability}
\newacronym{dit}{DIT}{data-operand independent timing}
\newacronym{gds}{GDS}{gather data sampling}
\newacronym{fpga}{FPGA}{field-programmable gate array}
\newacronym{got}{GOT}{global offset table}
\newacronym[longplural={instruction set architectures}]{isa}{ISA}{instruction-set architecture}
\newacronym{ir}{IR}{intermediate representation}
\newacronym{ip}{IP}{intellectual property}
\newacronym{jit}{JIT}{just-in-time}
\newacronym{gep}{\textsf{GEP}}{\textsf{GetElementPtr}}
\newacronym{ncsc}{NCSC}{National Cyber Security Centre}
\newacronym{mte}{MTE}{Memory Tagging Extension}
\newacronym{mrs}{MRS}{\texttt{malloc} revocation shim}
\newacronym[hyphenated={operating-system}]{os}{OS}{operating system}
\newacronym{pc}{PC}{program counter}
\newacronym{pcc}{\texttt{pcc}}{program counter capability}
\newacronym{rtl}{RTL}{register-transfer level}
\newacronym{rsb}{RSB}{return stack buffer}
\newacronym{sp}{SP}{stack pointer}
\newacronym{soc}{SoC}{system-on-chip}
\newacronym{stl}{STL}{store to load}
\newacronym{dma}{DMA}{direct memory access}
\newacronym{vcd}{VCD}{value change dump}
\newacronym{wns}{WNS}{worst negative slack}
\newcommand{\BLACKOUT}{\acrshort{blackout}\xspace}

\newcommand{\ba}{\glsentrydesc{ba}\xspace}

\newcommand{\bas}{\glsentrylongpl{ba}\xspace}
\newcommand{\bc}{\glsentrydesc{bc}\xspace}
\newcommand{\bcs}{\glsentrylongpl{bc}\xspace}
\newcommand{\BC}{\Glsentrydesc{bc}\xspace}
\newcommand{\BCs}{\Glsentrylongpl{bc}\xspace}
\newcommand{\bcadj}{\glshyph{bc}\xspace}
\newcommand{\Bd}{\Glsentrydesc{bd}\xspace}
\newcommand{\bd}{\glsentrydesc{bd}\xspace}

\newcommand{\bvs}{\glsentrylongpl{bv}\xspace}

\newcommand{\brs}{\glsentrylongpl{br}\xspace}
\newcommand{\blindedattribute}{\texttt{\_\_blinded}\xspace}
\newcommand{\annotatetype}{\texttt{annotate\_type}\xspace}

\newcommand{\blindedmalloc}{\texttt{blinded\_malloc}\xspace}

\newcommand{\OverheadRelativeToCheri}{$1.5\%$\xspace}
\newcommand{\OverheadRelativeTobaseline}{$23.5\%$\xspace}

\newcounter{tncnt}
\newcounter{mgcnt}
\newcounter{hecnt}
\newcounter{nacnt}

\begin{document}

\bstctlcite{ACMart:BSTcontrol}

\ifarxiv
\settopmatter{printacmref=false}
\renewcommand\footnotetextcopyrightpermission[1]{}
\pagestyle{plain} 
\fi

\ifnotanonymous
\StopCensoring
\fi

\censorruleheight=4ex
\title{\colorbox{black}{\texttt{\textcolor{white}{BLACKOUT}}}: Data-Oblivious Computation with Blinded Capabilities}
\renewcommand{\shorttitle}{\texttt{BLACKOUT}: Data-Oblivious Computation with Blinded Capabilities}

\settopmatter{authorsperrow=4}
\author{\censor{Hossam ElAtali}}
\ifnotanonymous
\authornote{Both authors contributed equally to this research.}
\fi
\affiliation{
  \institution{\censor{\textit{University of Waterloo}}}
  \ifnotanonymous
  \city{Waterloo}
  \country{Canada}
  \else
  \country{\censor{hossam.elatali@uwaterloo.ca}}
  \fi
}
\ifnotanonymous
\email{hossam.elatali@uwaterloo.ca}
\fi

\author{\censor{Merve Gülmez}}
\ifnotanonymous
\authornotemark[1]
\fi
\affiliation{
  \institution{\censor{\textit{Ericsson Security Research}}}
  \ifnotanonymous
  \city{Kista}
  \country{Sweden}
  \else
  \country{\censor{merve.gulmez@ericsson.com}}
  \fi
}
\ifnotanonymous
\email{merve.gulmez@ericsson.com}
\fi

\author{\censor{Thomas Nyman}}
\affiliation{
  \institution{\censor{\textit{Ericsson Product Security}}}
  \ifnotanonymous
  \city{Kista}
  \country{Sweden}
  \else
  \country{\censor{thomas.nyman@ericsson.com}}
  \fi
}
\ifnotanonymous
\email{thomas.nyman@ericsson.com}
\fi

\author{\censor{N. Asokan}}
\affiliation{
  \institution{\censor{\textit{University of Waterloo}}}
  \ifnotanonymous
  \city{Waterloo}
  \country{Canada}
  \else
  \country{\censor{asokan@acm.org}}
  \fi
}
\ifnotanonymous
\email{asokan@acm.org}
\fi
\ifanonymous
\renewcommand{\shortauthors}{Anon. Submission Id: 2275}
\global\def\authors{Anonymous Author(s)}
\acmBooktitle{second review cycle of CCS'25}
\acmConference[Second review cycle of CCS'25]{}{October 13--17, 2025}{Taipei, Taiwan}
\setcopyright{none}
\else
\copyrightyear{2025}
\acmYear{2025}
\setcopyright{cc}
\setcctype{by}
\acmConference[CCS '25]{Proceedings of the 2025 ACM SIGSAC Conference on Computer and Communications Security}{October 13--17, 2025}{Taipei, Taiwan}
\acmBooktitle{Proceedings of the 2025 ACM SIGSAC Conference on Computer and Communications Security (CCS '25), October 13--17, 2025, Taipei, Taiwan}\acmDOI{10.1145/3719027.3765169}
\acmISBN{979-8-4007-1525-9/2025/10}

\ifnotarxiv
\fi
\fi

\begin{abstract}
Lack of memory-safety and exposure to side channels are two prominent, persistent challenges for the secure implementation of software.
Memory-safe programming languages promise to significantly reduce the prevalence of memory-safety  bugs, but make it more difficult to implement side-channel-resistant code.
We aim to address both memory-safety and side-channel resistance by augmenting memory-safe hardware with the ability for data-oblivious programming.
We describe an extension to the \acrshort{cheri} capability architecture to provide \emph{\bcs} that allow data-oblivious computation to be carried out by userspace tasks. 
We also present \acrshort{blackout}, our realization of \bcs on a FPGA softcore based on the speculative out-of-order CHERI-Toooba processor and extend the CHERI-enabled Clang/LLVM compiler and the CheriBSD operating system with support for \bcs.
\BLACKOUT makes writing side-channel-resistant code easier by making non-data-oblivious operations via \bcs explicitly fault.
Through rigorous evaluation we show that \acrshort{blackout} ensures memory operated on through \bcs is securely allocated, used, and reclaimed and demonstrate that, in benchmarks comparable to those used by previous work, \BLACKOUT imposes only a small performance degradation (\OverheadRelativeToCheri{} geometric mean) compared to the baseline CHERI-Toooba processor.
\end{abstract}

\maketitle
\hypersetup{
   pdftitle={\shorttitle}
}
\censorruleheight=2ex

\section{Introduction}\label{sec:intro}

Weak memory-safety and side-channel leakage are two significant security challenges for security-critical software, such as  cryptographic libraries.
Lack of memory-safety in programming languages such as C and C++, is one of the oldest, most persistent problems in computer security.
The urgency of addressing the impact of memory-unsafe code at scale has grown under recent regulatory scrutiny~\cite{ONCD24} leading cybersecurity authorities like US \acrshort{cisa}~\cite{CISA23} and UK \acrshort{ncsc}~\cite{NCSC24} to advocate for memory-safe languages like Rust~\cite{Klabnik18}, and memory-safe hardware, such as \gls{cheri}~\cite{Watson19} as potential ways to eliminate entire classes of memory-safety vulnerabilities.
\ifnotabridged
Rust is poised to benefit from increased uptake thanks to the advocacy from regulators, but has not yet been proven in many industry sectors, and a complete rewrite of the vast body of existing C/C++ code in use today at a grand scale is economically impossible.
\fi

\emph{Constant-time code} provides security against timing side-channels by preventing attackers from inferring secret data by observing the timing of operations and instructions, cache effects, etc.
Writing constant-time code  by hand is hard, evident from the many flaws discovered in production side-channel-resistant code~\cite{Borrello21}.
Consequently, industry recommendations urge all but the most specialized developers to refrain from developing constant-time code~\cite{Intel19} because even slight mistakes can lead to exploitable side channels.
For example, \glspl{cpu} today do not treat control flow as a secret, thus allowing secret-dependent control-flow to leak sensitive information through the timing of operations or their effects on caches in a complex, hard-to-control, but observable way.
Optimizing compilers can transform code intended to be constant-time into \emph{functionally equivalent} (in terms of inputs and output) machine instructions but, for instance, branch on a secret value~\cite{Schneider24, Gerlach25}.
Similarly, hardware optimizations like speculative execution that are transparent to software can be exploited to leak secret data through existing side channels~\cite{Lipp18,Kocher19}.
Unlike memory-safety defects, which manifest as crashes or other faulty program behavior once a bug is triggered, flaws in constant-time programming \emph{fail silently} and might be detected only after successful exploitation, if at all. 

This problem is exacerbated in languages such as Rust which employ higher-level abstractions than C while still being compiled down to highly-optimized machine code.
Several unsuccessful attempts have been made to introduce constant-time-programming abstractions to the core Rust language~\cite{Klabnik15,Laeder16,Ben-Yehuda18}, leaving developers with few viable approaches for constant-time Rust programming.
Experts implementing Rust libraries (``crates'') providing primitives with constant-time properties acknowledge that such efforts are fundamentally limited because side-channel resistance is not a property of the software alone, but that of a system comprising both software and hardware~\cite{lovecruft25}.
A complete solution would require substantial changes to the compilers to allow severely restricting optimization on variables containing secret information, incurring significant cost to performance.
We discuss the alternatives employed in real-world cryptographic libraries in \Cref{sec:relatedwork}.

Memory-safe hardware, such as \gls{cheri},\ifnotabridged\xspace{} provides an alternative way to ensure memory-safety properties, particularly in systems where C and C++ will remain the dominant languages for the foreseeable future.
\gls{cheri}\fi{} effectively addresses memory safety, but does not inherently protect against side-channel attacks.
Previous work to harden \gls{cheri} against transient execution attacks~\cite{Fuchs24} does not protect against non-transient side-channel attacks.
On the other hand, inherent resistance against side channels would also be effective against transient execution attacks that employ such side channels for inferring their effects on microarchitectural properties like cache state.
In principle, \gls{cheri}-capable hardware could be complemented with existing data-oblivious computation solutions like OISA~\cite{Yu19} or BliMe~\cite{ElAtali24} that protect against side channels.
However, these approaches introduce considerable performance overhead because they require the addition of hardware-managed tag bits to memory.
\ifnotabridged
The adoption of hardware-accelerated memory tagging is hampered by significant meta-data storage and memory traffic overheads, since tags must be managed for  every piece of data in memory.
\fi

\noindent
\textbf{This paper and contributions.}
In this work, we propose \gls{blackout}, an extension to the \gls{cheri} memory-safety architecture. It provides novel \emph{\bcs} to augment \gls{cheri} with hardware-enforced taint tracking to guarantee confidentiality of secret data against conventional and speculative side channels.
Unlike prior approaches, \bcs avoid the need for additional tag bits in memory (beyond those employed by the baseline \gls{cheri} design).
Inside the \gls{cpu}, registers are extended with a \emph{blindedness} bit, which is set when the register is loaded with secret data.
\Gls{alu} operations incorporate taint tracking, propagating blindedness bits from operands to destination registers; therefore, any data derived from secret data is also marked as secret.
Instructions operating on such data in registers ensure that secret data is only written to memory through \bcs which have exclusive-access to ``blinded'' areas of memory.
Crucially, any attempt to misuse secret data\textemdash such as affecting control flow or as an address in memory operations\textemdash results in a fault, effectively preventing side-channel leakage.

We also present a \emph{data-oblivious programming model} for \gls{cheri} C which, aided by our \bcadj-aware Clang/LLVM compiler helps programmers in writing constant-time code for \BLACKOUT.
Our changes to the LLVM compiler infrastructure are \emph{non-invasive}, consisting of compiler passes that do not interfere with existing compiler optimizations, but instruct \BLACKOUT hardware about blinded variables.
\BLACKOUT allows developers to write constant-time code with minimal additional annotations, benefit from compile-time diagnostics, and turn previously silent constant-time bugs into explicit errors reported through \gls{cheri}'s exception mechanism.

To demonstrate the practicality of our approach, we realize \bcs on the CHERI-RISC-V architecture on a \gls{fpga} softcore based on the speculative out-of-order CHERI-Toooba processor and integrate \bcs into the CheriBSD \gls{os} and software stack.
We evaluate the area and performance overheads using data-oblivious benchmarks from seminal works in data-oblivious computing for comparability.
\BLACKOUT offers \textbf{the first unified solution to memory safety and side-channel confidentiality with minimal overhead}, addressing key limitations of existing methods. In summary, our contributions are:
\begin{itemize}[topsep=0pt]
    \item A hardware-software co-design for \emph{\bcs} that extends \gls{cheri} with the ability to carry out data-oblivious computation in userspace tasks (\Cref{sec:design}). 
    \item \BLACKOUT: a realization of \bcs on a CHERI-RISC-V  \gls{fpga} softcore based on the speculative out-of-order CHERI-Toooba processor \acrshort{ip} (\Cref{sec:hardware}).
    \item A programming model and software stack for \bcs and support for \BLACKOUT in the CHERI-enabled Clang/LLVM compiler and CheriBSD \gls{os}, thus enabling a broad class of applications to benefit from \BLACKOUT (\Cref{sec:sw-stack}).
    \item Evaluation of \BLACKOUT using the CoreMark industry-standard performance benchmark and data-oblivious benchmarks from prior work, showing minimal performance impact on data-oblivious code compared to the baseline CHERI-RISC-V processor (\OverheadRelativeToCheri{}), and moderate impact compared to a processor with neither \gls{cheri}'s memory-safety  nor \BLACKOUT enforcement (\OverheadRelativeTobaseline) (\Cref{sec:eval}).
\end{itemize}

Source code artifacts for the \BLACKOUT hardware and software stack are available at \url{https://github.com/blindedcapabilities}.

\section{Background}
\subsection{The \gls{cheri} capability architecture}\label{sec:cheri}

\begin{figure*}[t]
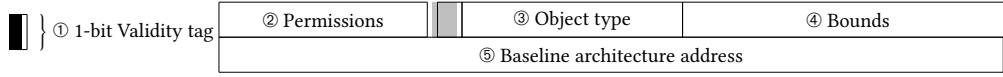

\begin{lrbox}{\bfbox}
    \begin{bytefield}[bitwidth=0.6em]{1}
        \begin{rightwordgroup}{\dCOne{} 1-bit Validity tag}
            \colorbitbox{black}{1}{\strut}
        \end{rightwordgroup} \\
    \end{bytefield}
    \begin{bytefield}[bitwidth=0.6em]{64}
        \bitbox{17}{\dCTwo{} Permissions} & \colorbitbox{lightgray}{2}{} &
        \bitbox{18}{\dCThree{} Object type} & \bitbox{27}{\dCFour{} Bounds} \\
        \bitbox{64}{\dCFive{} Baseline architecture address}
    \end{bytefield}%
\end{lrbox}
\resizebox{0.75\linewidth}{!}{\usebox{\bfbox}}
    \vspace{-0.2cm}
    \caption{In-memory representation of \gls{cheri} capabilities adapted from Watson et al.~\cite{Watson19}}\label{fig:chericap}
    \vspace{-0.2cm}
\end{figure*}

\gls{cheri} is an \gls{isa} extension that integrates a capability-based hardware-software co-design for memory protection.
It extends conventional \glspl{isa} with hardware-supported capabilities to verify memory accesses via code or data pointers.
The \gls{cheri} \gls{isa} specification~\cite{Watson23a} defines how capabilities are represented in registers and memory and offers capability-aware instructions to manipulate them. 
Current implementations of CHERI support the RISC-V and Arm8-A instruction sets with \gls{cheri}-enabled processors developed by Arm~\cite{Grisenthwaite22}, Microsoft~\cite{Amar23a} and companies and universities in the RISC-V ecosystem.

\Cref{fig:chericap} illustrates the in-memory representation of a \gls{cheri} capability. 
Each capability is double the width of the native pointer type: 128~bits on 64-bit platforms and 64~bits on 32-bit platforms.
A single-bit \emph{validity tag} \dCOne, stored separately, ensures integrity by invalidating capabilities manipulated by non-capability-aware instructions.
Capability-aware instructions preserve tags as long as the operation on a capability is valid but prevent unauthorized manipulation and injection of arbitrary capabilities.
These include the \gls{csc} and \gls{clc} instructions which are used to store, respectively load, capabilities to and from locations in memory identified by a capability operand.
In a \gls{cheri}-enhanced architecture, address and general-purpose \gls{cpu} registers are extended to store the full capability representation.
For example, in CHERI-RISC-V, all general-purpose registers are 128-bit \emph{capability registers}; the \gls{pc} and \gls{sp} are  represented by \gls{pcc} and \gls{csp} registers, respectively.
The capabilities themselves include several fields:
\begin{itemize}[topsep=0pt]
    \item \textbf{Permissions} (\dCTwo) define allowed operations.
    \item \textbf{Object type} (\dCThree) enables temporary ``sealing'', rendering a capability unusable until it is ``unsealed'' by a special instruction. This enables opaque pointer types and fine-grained in-process isolation.
    \item \textbf{Bounds} (\dCFour) specify the accessible memory range relative to the baseline architecture address (\dCFive). They are stored in a compressed format~\cite{Woodruff19} to reduce memory footprint, but require stricter alignment on larger object allocations.
\end{itemize}

New capabilities are always derived from an existing capability, with their lineage traceable to initial boot-time capabilities.
\gls{cheri} enforces \emph{monotonicity}  ensuring newly created capabilities cannot exceed the permissions or bounds of their parent. 
The \texttt{candperm} instruction allows the permissions of a capability to be dropped according to a given mask, but not gained.
Controlled exceptions, such as sealed capabilities for exception handling and compartmentalization, allow limited non-monotonicity.

 \paragraph{Spatial- and temporal-safety enforcement.}
The bounds, stored with the virtual address, underpin \gls{cheri}'s spatial memory safety.
Each memory allocation is associated with a capability describing its valid address range and access permissions.
This enables inherent spatial memory-safety properties. 
To provide temporary safety for heap-based allocations, CHERI requires a capability revocation mechanism, such as Cornucopia~\cite{Filardo24}.
Cornucopia scans for capabilities pointing to freed memory, allowing such stale capabilities to be revoked.
Extensions to the \gls{cheri} software and hardware have also explored sandboxing~\cite{Chisnall17}, and initialization safety~\cite{Georges21,Gulmez25}.

\paragraph{\gls{cheri}-enabled software stack.}
CheriBSD~\cite{Davis19} is a modified version of the open-source FreeBSD operating system, designed to support CHERI-RISC-V both in emulation and on hardware.
It is a fully functional \gls{os} prototype, demonstrating how \gls{cheri} support can be integrated into a conventional \gls{os} design.
The CheriBSD kernel and userspace can be built in ``\emph{pure-capability}'' \gls{cheri} C/C++ which means all conventional memory pointers are replaced by corresponding capabilities by the CHERI-LLVM compiler.
In pure-capability mode, compiler-generated code derives bounded capabilities for automatic (stack-allocated) variables from the \gls{csp}.
For global variables, the run-time linker populates into a capability-aware \gls{got}, also referred to as the \emph{captable}.
Capabilities for heap-allocated data is handled by Cornucopia which is integrated into CheriBSD via a \emph{shim layer}, known as the \gls{mrs}, which sits on top of the underlying BSD libc's memory allocator.
Consequently, CheriBSD enforces both spatial and temporal memory safety for \gls{cheri}-enabled software.

\subsection{Side-Channel Leakage}\label{bg:side-channels}

Side channels are unintended outputs of a system that can leak information about its behavior. CPU side channels are caused by implementation details, often the result of performance optimizations, such as caching mechanisms and speculative execution. Attackers can extract sensitive information by observing variations in execution time, power consumption, or electromagnetic emissions that are inadvertently output by these microarchitectural features. In this paper, we focus on timing side channels as they are the most commonly used in remote attacks.

A prominent example of timing side channels is cache timing, which leaks information by measuring cache access latency. By comparing the latency to a predetermined threshold, software can determine whether an address is already cached by the \gls{cpu}. Prime+Probe~\cite{Osvik06}, FLUSH+RELOAD~\cite{Yarom14} and Flush+Flush~\cite{Gruss16} are example side-channel attacks using different techniques to make this timing difference dependent on secret data and extract it from the system. Transient attacks like Spectre~\cite{Kocher19} and Meltdown~\cite{Lipp18}, expose new ways to access (architecturally inaccessible) secrets and leak them using side channels like cache timing.
Speculative taint-tracking techniques~\cite{Yu19a}, such as SpectreGuard~\cite{Fustos19}, ConTExT~\cite{Schwarz20}, and ProSpeCT~\cite{Daniel23}, mitigate Spectre-style attacks by delaying instructions dependent on speculatively loaded secrets, but mitigating side-channels in general falls on the developer. 

Preventing side-channel leakage is difficult for software developers. Software is often written in high-level languages that abstract away the individual \gls{cpu} instructions operating on secret data, leaving decisions to the underlying compilers and software stack. Even lower-level languages like C do not provide developers with sufficient controls to prevent side-channel leakage, as the compiler can introduce side channels in the generated assembly. Furthermore, even with seemingly side-channel-free assembly, hardware optimizations transparent to software (and often undocumented), such as speculation and out-of-order execution, can inadvertently leak secrets, as described above with the Spectre vulnerabilities. The challenge of preventing side-channel leakage is therefore present at every level of the software and hardware stack, and must be solved through hardware-software co-design.

\subsection{Data-Oblivious Instruction Set Architectures}\label{bg:data-oblivious-isa}

Data-oblivious computation refers to algorithms and software that process data in a manner where their control-flow and memory access patterns are not affected by their input data.
Constant-time code, which hardens software against side-channels, relies on data-oblivious algorithms and that the \gls{cpu} cycles consumed by individual hardware instructions are independent of their operands' values\footnotemark{}.
Historically, developers of security-critical code sensitive to side-channels, e.g., cryptographic implementations, have relied on obscure information on instruction timing differences obtained through extensive experiments on different \glspl{isa}~\cite{Pornin18a}.
In recent years, major processor manufacturers have began to document instructions with \gls{dit}~\cite{Intel23} and incorporate \gls{dit} modes into their designs~\cite{Intel23a, lowRISC25}.

The challenge of ensuring the correctness of data-oblivious software implementations, however, remains.
Verifying whether a program written in a high-level language is data-oblivious has a number of challenges (\Cref{sec:relatedwork}).
Chief among them is that data-obliviousness can only be defined at a machine-code level.
This observation has motivated the creation of \emph{data-oblivious \glspl{isa}} that, similar to data operand independent timing modes, provide modes of operation for \glspl{cpu} to ensure that instructions operating on sensitive data do so in a data-oblivious manner.
Previous works use hardware-enforced taint tracking using ``\emph{blindedness tags}''~\cite{ElAtali24a} and dedicated memory partitions~\cite{Yu19}.
These approaches result in best-case performance overheads from 8\% to 35\% (worst case involving an order of magnitude slowdown).
Memory tagging exhibits poor performance as the number of tag bits increase. 
Schemes such as the Arm architecture's \gls{mte}, used for memory-safety sanitation, limit the number of tag bits to 4.
\gls{mte} further mitigates performance impact by allowing tag checks to occur asynchronously.
But asynchronous tag checks are not practical for maintaining data-obliviousness as data-oblivious \glspl{isa} must be able to prevent data accesses before confidentiality is violated.

\footnotetext{``Constant-time code'' is a misnomer, as even side-channel-resistant code may still exhibit variable latency if the latency differences are not due to the (secret) data values. Common sources of such latencies include instruction-level latencies due to frequency scaling, or unpredictable contention of data buses for loads and stores.}

\begin{figure*}[t]
    \includegraphics[width=\textwidth]{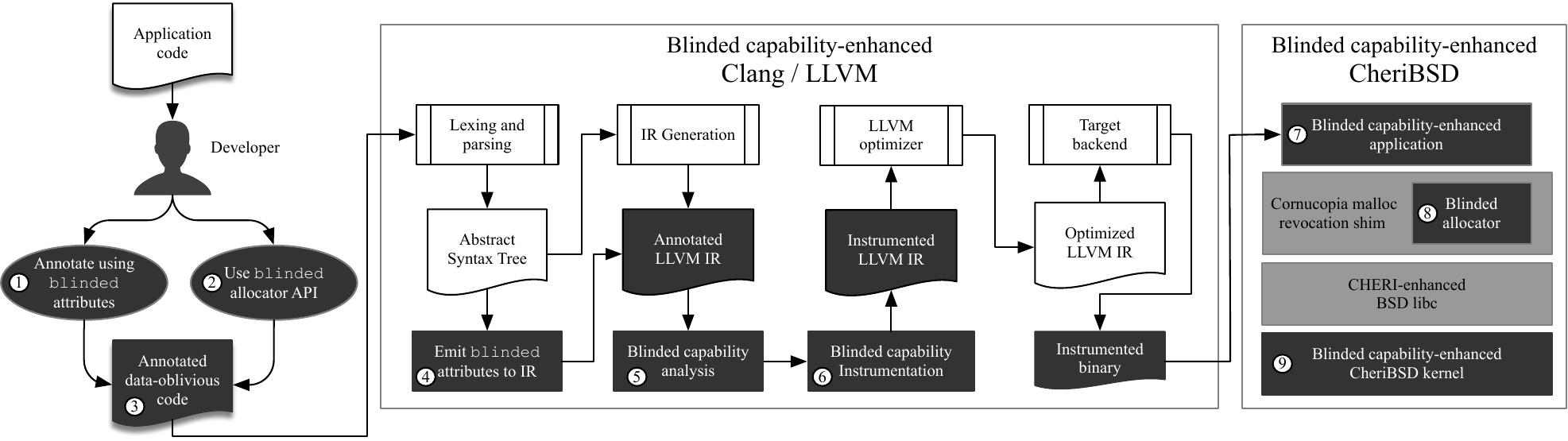}
    \vspace{-0.4cm}
    \caption{High-level overview of the components of the \bc software stack. Dark gray denotes additions or modifications needed to support \bcs, while light grey indicates components inherited from the \gls{cheri} architecture. White denotes components without significant changes.}\label{fig:swarch}
    \vspace{-0.2cm}
\end{figure*}

\section{System and Adversary Model}
We assume the same system model as in CHERI. We trust the \gls{os} to prevent memory containing sensitive information from being exposed to other processes after a process exits, either normally or as a result of \gls{cheri} exceptions. We assume memory-safety properties provided by \gls{cheri}, which prevent adversaries from tampering with the program's control flow or directly inferring memory contents beyond that which are exposed during normal program operation.
We also assume a \gls{dit}~\cite{lowRISC25} mode is available.

We assume the same adversary model as in CHERI. In addition, we assume the adversary has the ability to observe side channels during a program's execution, possibly from another process running on the system simultaneously. In this work, we consider side-channel threats that can be mitigated by data-oblivious software, including timing side channels, such as cache timing~\cite{Osvik06,Yarom14,Gruss16,Chiang25}, instruction timing~\cite{Pornin18} (with DIT) and port-contention timing~\cite{Aldaya19}. We also consider transient execution attacks that leak information \emph{loaded} by mis-speculated instructions within the same address space, such as Spectre~\cite{Kocher19,Maisuradze18,Koruyeh18,Horn18}, in scope, even if mis-training can be done across address spaces~\cite{Canella19a}. Attacks with transient instructions that can load data \emph{across} address spaces, such as Meltdown and microarchitectural data sampling (including load-store-buffers)~\cite{Canella19}, and transient capability forgery, such as Meltdown-CF~\cite{Fuchs24}, are out of scope, but can be prevented by using \glspl{capSpecCon}~\cite{Fuchs24}.
Attacks that require physical access to the system, e.g., to measure power consumption, or those through \gls{dma} peripherals are out of scope.

\section{Goals and Challenges}\label{sec:goalsandchallenges}

In this section, we outline our overarching goals and the primary challenges associated with them.

\subsection{Goals and Primary Challenges}\label{sec:goals}

\paragraph{Adapting hardware-enforced taint tracking to capability-based architectures.}
Our primary goal is integrating data-oblivious computation with the \gls{cheri} protection model.
While it may seem straightforward to implement a data-oblivious \gls{isa} on top of an \gls{cheri}-enabled architecture, there are two key challenges to overcome.
First, the \gls{cheri} architecture is intended to be applied onto conventional \glspl{isa} such as RISC-V and Arm.
Bolting an existing data-oblivious \gls{isa} (\Cref{bg:data-oblivious-isa}) on top of \gls{cheri} will require invasive changes to \gls{cheri} and/or the underlying architecture making real-world adoption less realistic.
Second, significant changes to the architecture will negatively impact performance.

To address these challenges, we propose enhancing the existing \gls{cheri} capability model with \emph{\bcs} which ensure data accessed through them is operated on exclusively in a data-oblivious manner.
Memory managed by \bcs is referred to as \emph{\bd}.
This avoids the need to propagate blindedness tags to memory thus overcoming the drawbacks of simply integrating an existing data-oblivious \gls{isa} with \gls{cheri}~\cite{ElAtali24,Yu19}.

\paragraph{Practical programming model for \bcs.}
Our second goal is introducing a practical programming model for \bcs.
\gls{cheri}-enabled languages (CHERI C/C++~\cite{Watson20}) enforce memory-safety properties (\Cref{sec:cheri}) at run-time whereas data-oblivious \glspl{isa} enforce oblivious access to secret data (\Cref{sec:design}).
Thus, our model must adhere to both \gls{cheri} and data-oblivious properties.

A particular challenge arises because the \gls{cheri} compiler occasionally allows memory accesses without explicit capabilities (e.g., certain stack accesses).
To ensure such accesses do not target \bd (which would raise a hardware fault), our \bc-enhanced compiler must explicitly enforce that all accesses to blinded stack variables occurs through \bcs.
We describe this compiler enhancement further in \Cref{sec:design} and \Cref{sec:sw-stack}.

\subsection{Additional Challenges}\label{sec:challenges}

Integrating blinded memory management with the \gls{cheri} protection model introduces several additional challenges.

\paragraph{Maintaining capability monotonicity.} 
Introducing \bcs adds new privileges that must align with \gls{cheri}'s existing monotonicity properties.
Our design considers ``non-blinded'' a distinct permission, which, when removed, permanently classifies a capability as a \bc.
This upholds monotonicity\textemdash\bcs cannot be promoted to regular, non-blinded capabilities.

\paragraph{Exclusive access to blinded memory.}
As \bcs{} are based on \gls{cheri} capabilities, they inherently face the \emph{revocation problem}.
Deallocated memory regions may still be accessible through residual capabilities, risking use-after-free vulnerabilities and inadvertent disclosure of blinded data.
Our design addresses this by ensuring new \bcs obtain exclusive access to newly allocated \bd inside the process.
Thus, \bd cannot be subject to use-after-free conditions from residual, non-blinded capabilities to it.

\paragraph{Securely reclaiming blinded memory.}
\Bd must be securely reclaimed, as residual sensitive data may remain upon deallocation.
Since \gls{cheri} does not inherently guarantee memory initialization safety, we need to ensure secure deallocation through automatic memory clearing with \bc-enhanced compiler and a \bc-enhanced allocator (\Cref{sec:sw-stack}).

\begin{figure*}[t]
\begin{centering}
    \includegraphics[width=1.4\columnwidth]{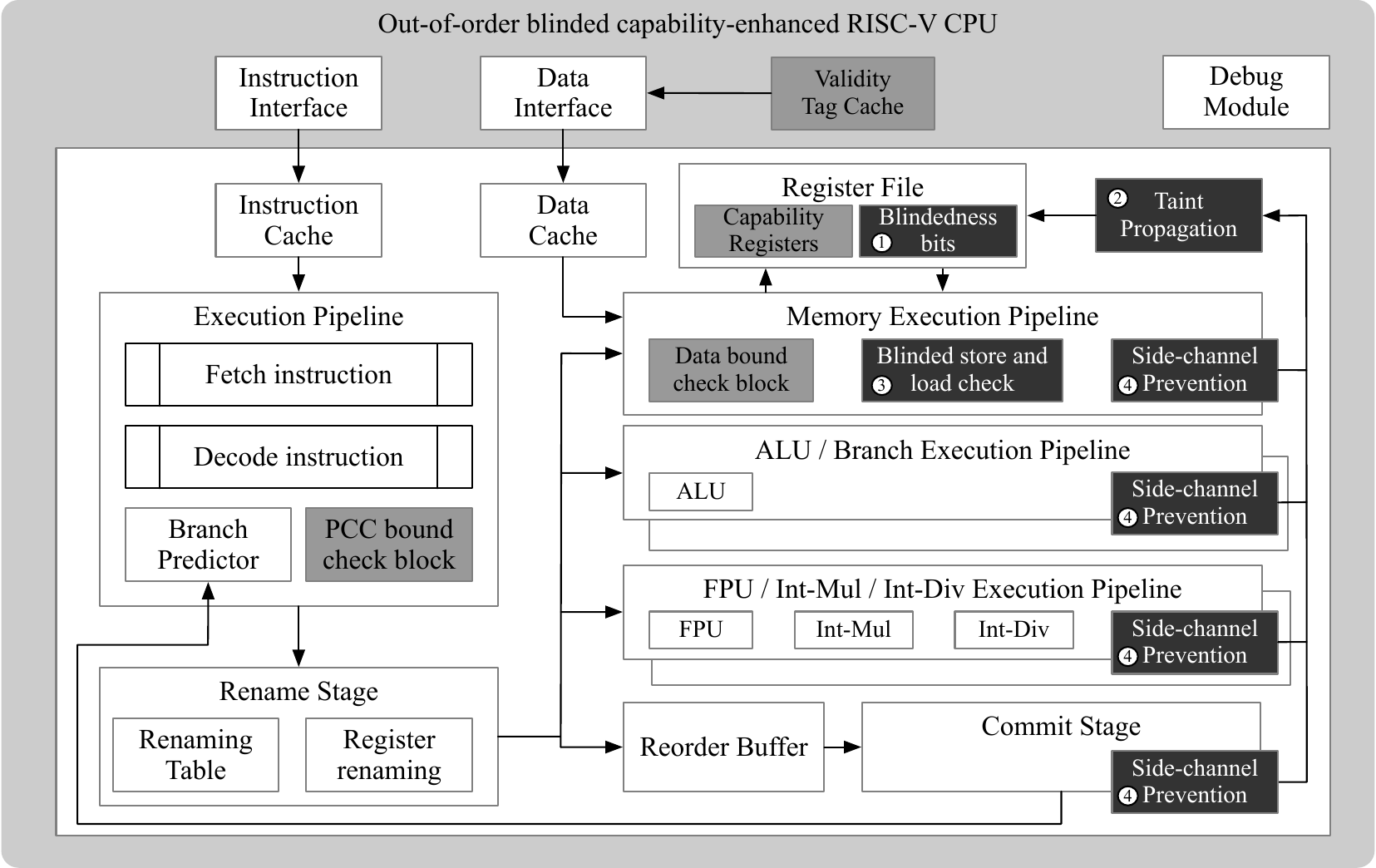}
    \vspace{-0.2cm}
    \caption{High-level overview of the \bc-enhanced CHERI-RISC-V \gls{cpu} hardware components. Dark grey denotes additions or modifications needed to support taint tracking, while light grey indicates additions common to \gls{cheri}-enabled \glspl{cpu}. White denotes components without significant changes.}\label{fig:hwarch}
\end{centering}
\vspace{-0.2cm}
\end{figure*}

\section{Blinded Capability Design}\label{sec:design}
\BCs introduce a new programming model that developers must follow to protect their sensitive data.
While providing a completely unrestricted programming model can seem appealing, it 
\begin{inparaenum}[1)]
  \item cannot guarantee exclusive access to blinded data (\Cref{sec:challenges}) and 
  \item does not guide the developer towards avoiding hardware faults caused by security violations.
\end{inparaenum}
The end result is that the developer must manually analyze and transform their code to prevent it from faulting.
The goal of our programming model is therefore to provide strong security guarantees for memory safety and side-channel protection by guiding the developer using compiler warnings/errors, and, where possible, automatically applying code transformations.

\subsection{Software Architecture}

\Cref{fig:swarch} shows a high-level overview of the \bc software stack.
Developers can indicate to the compiler that a variable is sensitive by annotating its declaration with the \ba{}~\dCOne.
We refer to such variables as \emph{\bvs}.
 The compiler will then ensure that any accesses to \bvs are permitted only through \bcs, guaranteeing that the values of these variables will always be blinded when loaded into registers.
 We refer to registers holding blinded data as \emph{\brs}.
 \Bd can be dynamically allocated using the \texttt{blinded} allocator \gls{api}~\dCTwo which returns a \bc.
 Similarly, the compiler enforces that references to \bd cannot be assigned to non-blinded variables. 
Inside the compiler, we modify the Clang front-end to emit \bas to the LLVM \gls{ir} of the compiled program \dCFour.
These are used by the \bc analysis~\dCFive and instrumentation passes~\dCSix to produce a \bc-instrumented application binary.
Global variables annotated with \ba{} are reserved from a separate \texttt{blinded data} section by the static linker.

At run-time, the \bcadj-enhanced application~\dCSeven requires platform support in the form of the blinded allocator~\dCEight and an \gls{os} with a \gls{cheri}-capable kernel that has been altered to ensure its internal handling of capabilities does not trigger \bc violations.
C startup code added by the compiler to \dCSeven is responsible for initializing capabilities pointing to data in the blinded data section as \bcs.
System software not making use of \bcs does not require modification (apart from their adaptation to \gls{cheri}).
We discuss the compiler enhancements and operating system changes in detail in \Cref{sec:sw-stack}.

\begin{figure*}[t]
\noindent\begin{minipage}{\columnwidth}
\begin{adjustbox}{width=\textwidth,keepaspectratio}
\begin{lstlisting}[style=CStyle, label={lst:oblivious_select}, caption={Data oblivious conditional select function which returns one of the arguments \texttt{x} or \texttt{y} based on the value of \texttt{cond} demonstrating the inference capabilities of the \bc-enhanced compiler.}]
#define __blinded [[clang::annotate_type("blinded")]]

%*\dOne\hspace{-.5em}*) __attribute__ ((blinded))
int data_oblivious_select(bool cond, int x, int y) {
    %*\tikzmark{abegin}*)
    bool __blinded c; // c declared blinded and
                      // accessed via blinded capability
    %*\tikzmark{aend-bbegin}*)
    int res;        // res not declared blinded
                    // but blindedness is inferred
    %*\tikzmark{bend}*) %*\tikzmark{bcomment}*)
    %*\tikzmark{cbegin}*)
    c = cond;   // Uses store via blinded capability
                // (argument already in register)
    %*\tikzmark{cend}*) %*\tikzmark{ccomment}*)
    %*\tikzmark{dbegin}*)
    res = (x * c) + (y * (!c)); // HW propagates
                                // blindedness to res
    %*\tikzmark{dend}*) %*\tikzmark{dcomment}*)
    %*\tikzmark{ebegin}*)
    return res;
}   %*\tikzmark{eend}*) %*\tikzmark{ecomment}*)

\end{lstlisting}
\begin{tikzpicture}[remember picture, overlay, thick]
    \draw[decorate,decoration={brace,amplitude=5pt,mirror}] ([shift={(-4pt,-2pt)}]pic cs:abegin) 
    -- ([shift={(-4pt,5pt)}]pic cs:aend-bbegin) 
    coordinate[midway,xshift=-6pt](Btip);
    \draw[rounded corners] (Btip) -- +(0,0)  node[left,none]
    {\scriptsize \dTwo};

    \draw[decorate,decoration={brace,amplitude=5pt,mirror}] ([shift={(-4pt,-2pt)}]pic cs:aend-bbegin) 
    -- ([shift={(-4pt,5pt)}]pic cs:bend) 
    coordinate[midway,xshift=-6pt](Btip);
    \draw[->, rounded corners] (Btip) -- +(-3pt,0) |- (pic cs:bcomment) node[right,comment,thin]
    {\scriptsize \dThree Compiler infers \texttt{res} blinded from \dFive};

    \draw[decorate,decoration={brace,amplitude=5pt,mirror}] ([shift={(-4pt,-2pt)}]pic cs:cbegin) 
    -- ([shift={(-4pt,5pt)}]pic cs:cend) 
    coordinate[midway,xshift=-6pt](Btip);
    \draw[->, rounded corners] (Btip) -- +(-3pt,0) |- (pic cs:ccomment) node[right,comment,thin]
    {\scriptsize \dFour Compiler knows \texttt{c} is already blinded based on declaration};

    \draw[decorate,decoration={brace,amplitude=2pt,mirror}] ([shift={(-4pt,-2pt)}]pic cs:dbegin) 
    -- ([shift={(-4pt,5pt)}]pic cs:dend) 
    coordinate[midway,xshift=-6pt](Btip);
    \draw[->, rounded corners] (Btip) -- +(-3pt,0) |- (pic cs:dcomment) node[right,comment,thin]
    {\scriptsize \dFive Compiler infers \texttt{res} is blinded from this assignment};

\end{tikzpicture}
\end{adjustbox}
\end{minipage}
\hfill
\begin{minipage}{\columnwidth}
\begin{adjustbox}{width=.95\textwidth,keepaspectratio}
\begin{lstlisting}[style=CStyle, label={lst:bad_func}, caption={Non-data-oblivious function which attempts to branch on a blinded condition and write blinded data to a possibly non-blinded output parameter. This code is rejected by the blinded capability-enhanced compiler.}]
void bad_func(bool cond, int x, int *out) {
    %*\tikzmark{fbegin}*)
    int __blinded a = x;   // a declared blinded
    int b; 
    %*\tikzmark{fend}*)
    %*\tikzmark{gbegin}*)
    if(cond)
        b = a;   // HW propagates blindedness to b
    %*\tikzmark{gend}*) %*\tikzmark{gcomment}*)
    %*\tikzmark{hbegin}*)
    *out = a;    // HW fault if out non-blinded (%*\color{green!50!black}{\labelcref{req:blindedstore}}*))
    %*\tikzmark{hend}*) %*\tikzmark{hcomment}*)
    %*\tikzmark{ibegin}*)
    if(a != 0)   // Violates %*\color{green!50!black}{\labelcref{req:controlflow}}*)
        b = a;
    else
        return
    %*\tikzmark{iend}*) %*\tikzmark{icomment}*)
    %*\tikzmark{jbegin}*)
    if (b != 0)     // HW fault if b blinded (%*\color{green!50!black}{\labelcref{req:controlflow}}*))
        *out = a;   // HW fault if out non-blinded (%*\color{green!50!black}{\labelcref{req:blindedstore}}*))
}   %*\tikzmark{jend}*) %*\tikzmark{jcomment}*)
\end{lstlisting}
\begin{tikzpicture}[remember picture, overlay, thick]
    \draw[decorate,decoration={brace,amplitude=5pt,mirror}] ([shift={(-4pt,-2pt)}]pic cs:fbegin) 
    -- ([shift={(-4pt,5pt)}]pic cs:fend) 
    coordinate[midway,xshift=-6pt](Btip);
    \draw[rounded corners] (Btip) -- +(0,0)  node[left,none]
    {\scriptsize \dCOne};

    \draw[decorate,decoration={brace,amplitude=5pt,mirror}] ([shift={(-4pt,-2pt)}]pic cs:gbegin) 
    -- ([shift={(-4pt,5pt)}]pic cs:gend) 
    coordinate[midway,xshift=-6pt](Btip);
    \draw[->, rounded corners] (Btip) -- +(-3pt,0) |- (pic cs:gcomment) node[right,comment,thin]
    {\scriptsize \dCTwo Compiler undecided on whether \texttt{b} becomes blinded};

    \draw[decorate,decoration={brace,amplitude=5pt,mirror}] ([shift={(-4pt,-2pt)}]pic cs:hbegin) 
    -- ([shift={(-4pt,5pt)}]pic cs:hend) 
    coordinate[midway,xshift=-6pt](Btip);
    \draw[->, rounded corners] (Btip) -- +(-3pt,0) |- (pic cs:hcomment) node[right,comment,thin]
    {\scriptsize \dCThree Allowed by compiler as \texttt{out} might be blinded};

    \draw[decorate,decoration={brace,amplitude=5pt,mirror}] ([shift={(-4pt,-2pt)}]pic cs:ibegin) 
    -- ([shift={(-4pt,5pt)}]pic cs:iend) 
    coordinate[midway,xshift=-6pt](Btip);
    \draw[->, rounded corners] (Btip) -- +(-3pt,0) |- (pic cs:icomment) node[right,comment,thin]
    {\scriptsize \dCFour Rejected by compiler as blinded \texttt{a} used in control-flow decision};

        \draw[decorate,decoration={brace,amplitude=5pt,mirror}] ([shift={(-4pt,-2pt)}]pic cs:jbegin) 
    -- ([shift={(-4pt,5pt)}]pic cs:jend) 
    coordinate[midway,xshift=-6pt](Btip);
    \draw[->, rounded corners] (Btip) -- +(-3pt,0) |- (pic cs:jcomment) node[right,comment,thin]
    {\scriptsize \dCFive Control-flow allowed by compiler as \texttt{b} might be non-blinded};
\end{tikzpicture}
\end{adjustbox}
\end{minipage}
\end{figure*}

\begin{table}
    \caption{Blindedness bit propagation and side-channel prevention rules enforced by \BLACKOUT hardware. x represents a ``don't care'' condition, which indicates the actual signal value or values have no impact on the decision outcome.}\label{tab:propagation-rules}
    \vspace{-0.3cm}
    \centering
     \resizebox{\columnwidth}{!}{
                \begin{tabular}{@{}cccccc@{}}
            \toprule
            \textbf{Instruction} & \textbf{Blinded} & \multicolumn{2}{c}{\centering \textbf{Blindedness}} & \textbf{Decision} & \textbf{Result} \\
            & \textbf{capability} & & & & \textbf{blindedness}\\
            \midrule
            \Large \textbf{Arithmetic / } & \multirow{2}{2.5cm}{\centering \Large -} & \textbf{Op1} & \textbf{Op2} & & \\
            \cmidrule{3-4}
             \Large \textbf{Logic} &  & \Large $a$ & \Large $b$ & \Large Propagate & \Large $a \vee b$ \\
            \midrule
            \multirow{5}{*}{\centering \Large \textbf{Branching}} & \textbf{in Addr Reg} & \textbf{Addr Reg} & \textbf{Condition Ops} & & \\
            \cmidrule{3-4}
             & \Large no  & \Large 0 & \Large 0 & \Large Allow & \Large - \\
             & \Large no  & \Large 1 & \Large x & \Large Fault & \Large - \\
             & \Large no  & \Large x & \Large 1 & \Large Fault & \Large - \\
             & \Large yes & \Large x & \Large x & \Large Fault & \Large - \\
            \midrule
            \multirow{4}{*}{\centering \Large \textbf{Load}} & \textbf{in Addr Reg} & \multicolumn{2}{c}{\textbf{Addr Reg}} & & \\
            \cmidrule{3-4}
             & \Large no         & \multicolumn{2}{c}{\Large 0} & \Large Propagate & \Large 0 \\
             & \Large yes        & \multicolumn{2}{c}{\Large 0} & \Large Propagate & \large 1 \\
             & \Large x & \multicolumn{2}{c}{\Large 1} & \Large Fault     & \Large - \\
            \midrule
            \multirow{6}{*}{\centering \Large \textbf{Store}} & \textbf{in Addr Reg} & \textbf{Addr Reg} & \textbf{Data Reg} & & \\
            \cmidrule{3-4}
             & \Large no  & \Large 0  & \Large 0 & \Large Allow & \Large - \\
             & \Large no  & \Large 0  & \Large 1 & \Large Fault & \Large - \\
             & \Large yes & \Large 0  & \Large x & \Large Allow & \Large - \\
             & \Large x   & \Large 1  & \Large x & \Large Fault & \Large - \\
            \bottomrule
        \end{tabular}}
        \vspace{-1em}
\end{table}

\subsection{Hardware architecture}\label{sec:design-hw-arch}

\Cref{fig:hwarch} shows the high-level overview of the hardware changes needed to support \bcs (shown in dark gray) applied to an out-of-order RISC-V core, such as MIT's RISCY-OO~\cite{Zhang25} or Bluespec's Toooba~\cite{Nikhil25} processor \gls{ip}.
Apart from the changes introduced by \gls{cheri} (shown in light gray) the majority of the changes necessary in the \gls{cpu} are to facilitate in-\gls{cpu} taint tracking.
In the register file, the general-purpose capability registers are extended to hold an additional \emph{blindedness bit}~\dCOne which tracks whether the register is blinded.
The taint-propagation hardware logic \dCTwo will in turn ensure that any data derived from these \brs, e.g., through arithmetic operations, will also ``be blinded'', i.e., it sets the blindedness bit of the destination register. 
 
 All blinded data is bound by a set of invariants (\labelcref{req:blindedstore,req:noblindedcap,req:overlap,req:controlflow,req:loadstore}), stemming from the need to enforce exclusive access to \bd (\Cref{sec:challenges}), and side-channel prevention.

\paragraph{Exclusive access invariants:}
\begin{enumerate}[label=I\arabic*, topsep=0pt, leftmargin=14pt]
    \item \label{req:blindedstore}Blinded data cannot be stored into memory using non-blinded capabilities.
    \item \label{req:noblindedcap}No capability of any kind can be stored into \bd.
    \item \label{req:overlap}Bounds of valid blinded and non-blinded capabilities must not simultaneously overlap. However, once a region is released (i.e., is no longer accessible through a valid capability, e.g., due to heap deallocations or stack frame pops), it can be assigned to future blinded or non-blinded capabilities.
\end{enumerate}

\paragraph{Side-channel protection invariants:}
\begin{enumerate}[label=I\arabic*, topsep=0pt, leftmargin=14pt]
\setcounter{enumi}{3}
    \item \label{req:controlflow}Control-flow instructions cannot use blinded operands as conditions or target addresses.
    \item \label{req:loadstore}Load and store instructions cannot use blinded operands as addresses.
\end{enumerate}

\labelcref{req:blindedstore} is enforced by hardware through additional \emph{blinded store and load checks}~\dCThree in the memory execution pipeline which is responsible for the execution of any load and store instructions, whereas \labelcref{req:controlflow,req:loadstore} are enforced by \emph{side-channel prevention} logic integrated into any pipeline which is responsible for instructions with blinded operands~\dCFour.
Any attempt by software to violate \labelcref{req:blindedstore,req:controlflow,req:loadstore} will result in a hardware fault.
\Cref{tab:propagation-rules} shows the full blindedness bit propagation and side-channel prevention rules enforced by \BLACKOUT.

\labelcref{req:overlap} cannot be efficiently enforced by hardware alone.
As mentioned in \Cref{sec:challenges}, blinded memory reclaimed when blinded capabilities are destroyed must be erased to avoid leaking blinded data; this is handled by the compiler and memory allocator for stack and heap variables, respectively (\Cref{fig:swarch}, \Cref{sec:sw-stack}).
The combination of hardware, compiler and software stack guarantees data confidentiality.

Data-oblivious computation ensures that control flow and memory accesses don’t depend on secret data.
Outputs, however, may be non-secret (e.g., decrypted data meant for users).
In practice, such output data cannot be left blinded indefinitely as some result of data-oblivious computation must eventually be extracted.
In \BLACKOUT, results can be marked non-secret by issuing non-blinded capabilities for result buffers.
This reveals the final, non-secret result of blinded computation by relaxing invariant \labelcref{req:overlap} in a controlled way to unblind the result.
For example, since \labelcref{req:overlap} is (in part) enforced by the blinded allocator, the allocator can expose a separate \gls{api} for obtaining blinded results that, in contrast to, regular \bd returns two capabilities for a dedicated area for blinded results: one which is blinded, and used to write to the area, and the other non-blinded, but only usable for reading the result.
However, the developer must use the \bc to the result area carefully to prevent accidental leakage of secret data. We discuss additional options to secure extraction further in \Cref{sec:capescrow}.

\subsection{Motivating Examples}

The concrete examples in \Cref{lst:oblivious_select,lst:bad_func} demonstrate how the hardware and the compiler work together to prevent \bd leakage.
The hardware ensures that taint tracking is both accurate and precise (no under- or over-tainting occurs in the hardware), and that attempted leaks cause a fault.
The compiler ensures that capabilities respect exclusive \bd access, and that reclaimed blinded stack memory is zeroed out.

The compile-time analysis is dependent on the developer providing \emph(initial) information about which variables are expected to contain sensitive information through \bas (\blindedattribute, \dTwo in \Cref{lst:oblivious_select}).
Optionally, the developer can also declare functions blinded~(\texttt{\_\_attribute\_\_((blinded))}, \dOne in \Cref{lst:oblivious_select}). This clearly identifies functions expected to return blinded data, improving developer ergonomics by enhancing the readability of data-oblivious code.
The compile-time analysis is not limited to the information provided by the developer.
For example, in \Cref{lst:oblivious_select} \dThree, the compiler can infer the variable \texttt{res} must be blinded due to the assignment from \texttt{c} tainting it at \dFive.

\Cref{lst:bad_func} shows a non-data-oblivious function which would fault at run-time due to violating \labelcref{req:controlflow} (and possibly \labelcref{req:blindedstore}).
This example demonstrates that the compiler can reject some non-data-oblivious programs outright.
However, as discussed in \Cref{bg:data-oblivious-isa}, compile-time data-obliviousness analysis cannot be sound as data-obliviousness is ultimately a property at machine-code level.
For example, at \dCTwo, the compiler cannot know whether \texttt{cond==1}. Thus, it \emph{cannot prove} that \texttt{b} is always blinded and must allow the control flow at \dCFive.
However, if \texttt{cond==1} at run time, the hardware will propagate the blindedness bit to \texttt{b} at \dCTwo; the subsequent branch at \dCFive will fault.

For \dCThree, the compiler does not know whether \texttt{out} is a blinded capability and must also allow compilation. If \texttt{out} at run time is referenced by an unblinded capability, the hardware will detect the violating store and raise a fault.

The hardware would fault at \dCFour due to violation of \labelcref{req:controlflow}. However, since it is detectable by the compiler, we force a compilation error to inform the developer early in the development cycle.
This is especially useful for code paths that are rarely exercised with real workloads and require fuzzing techniques to uncover. More annotations allow the compiler to make more informed decisions and detect violating operations that it cannot infer on its own.

\section{\BLACKOUT Blinded Capability Implementation}

We now describe \gls{blackout}, our realization of \bcs for CHERI-RISCV, CHERI-LLVM, and the CheriBSD \gls{os}.

\subsection{\BLACKOUT CHERI-RISC-V \gls{cpu}}\label{sec:hardware}
We implement \bcs in \gls{rtl} on CHERI-Toooba~\cite{Fuchs24}, a \gls{cheri}-enabled RISC-V processor based on Bluespec's speculative out-of-order Toooba \gls{ip}~\cite{Nikhil25}.
Unlike previous data-oblivious \glspl{isa}~\cite{Yu19,ElAtali24}, \BLACKOUT \emph{does not add any additional instructions to the underlying \gls{isa}}; all architectural changes are achieved by adjusting the \gls{cheri} permission model.

\paragraph{Non-oblivious access permission.}
We introduce a new ``non-oblivious access'' permission bit in the previously unused section of the CHERI capability representation.
The CHERI-Toooba IP allocates 12 bits for hardware permissions and 4 bits for user permissions, leaving 3 bits unused.
By default, the \bc bit in newly created capabilities is set to 1.
In this initial configuration, the enforcement of \bcs is turned off, allowing the capability to be used freely within its defined bounds.
When the existing \gls{cheri} \texttt{candperm} instruction (\Cref{sec:cheri}) is invoked with the \bc permission as an operand, it changes the non-oblivious access bit to 0, thereby enabling the enforcement of \bcs.
Defining the non-oblivious access bit this way means that a capability with this permission is \emph{allowed} to handle data in a non-oblivious manner.

\paragraph{Registers \& taint-tracking.}
CPU registers are extended with an additional blindedness bit to signify whether the data stored in the register is blinded. Any load instructions executed with a \bc set the blindedness bit of the target register to 1. Store instructions with a \bc only store the data in memory, but not the blindedness bit. Data confidentiality is maintained by the exclusive access invariants (\labelcref{req:blindedstore,req:noblindedcap,req:overlap}, \Cref{sec:design}), which ensure that blinded data stored in memory is only accessible through \bcs. Register-to-register instructions, such as arithmetic and logical operations, propagate the blindedness bit: if the output depends on a blinded input, the output becomes blinded. Overwriting a register containing blinded data with non-blinded data will result in that register's blindedness bit to be unset.

\paragraph{Limiting impact on capability-modifying instructions.}
We prohibit valid capabilities from being blinded (\labelcref{req:noblindedcap}, \Cref{sec:design}). The execution of any capability-modifying instruction with operands that would result in a blinded capability causes a fault. Enforcing this invariant allows us to avoid unnecessarily making capability-modifying instructions data-oblivious. Note that this enforcement does not break the \gls{cheri} programming model; ``\emph{blinded blinded capabilities}'' are themselves redundant as they can never be dereferenced to access memory in order to maintain confidentiality (\labelcref{req:loadstore}, \Cref{sec:design}).

\paragraph{Spilling registers containing blinded data.}
Blinded data can reside in any general-purpose capability register, including those designated as caller- or callee-saved by the RISC-V \gls{abi}. 
The CHERI-RISC-V compiler can generate code that spills \brs using \gls{csc} instructions, and restore them with \gls{clc} instructions.
Normally, spilling blinded registers onto the stack via the non-blinded \gls{csp} violates invariant \labelcref{req:blindedstore} (\Cref{sec:design}) as spills target non-\bd.
However, since stack memory allocated for registers spills is inaccessible by other capabilities than the \gls{csp}, \gls{cheri}'s memory-safety guarantees ensure that spilled register cannot be accessed improperly.
Thus, we permit \BLACKOUT's \gls{csc} and \gls{clc} instructions to spill and restore \brs \gls{csp}.

However, another challenge arises: the non-blinded \gls{csp} generates non-blinded data when storing spilled values.
This causes ambiguity when restoring spilled registers, as the \gls{cpu} cannot distinguish between blinded and non-blinded register spills.
To address this, \BLACKOUT{}'s \gls{csc} emits special \emph{\glspl{brr}} as a result of spilling \brs.
A \gls{brr} is a 128-bit structure containing the spilled value and 64-bit marker. 
This marker differentiates \glspl{brr} from conventional \gls{cheri} capabilities and maintains their validity tags.
When a spilled capability register is restored, the validity tag signals that the value restored from memory is either a capability or a \gls{brr}.
The marker indicates whether the value is a \gls{brr}, ensuring the destination register is correctly blinded.

\paragraph{Transient execution.}
As mentioned in \Cref{bg:side-channels}, hardware optimizations, such as speculation, can inadvertently introduce side channels when operating on secrets. To protect blinded data against such leakage, we ensure that any blinded data flowing to \emph{decision-making} parts of the hardware (such as branch predictors) are zeroed out. This guarantees that execution is truly oblivious to blinded data, both architecturally and transiently. This is similar to approaches used in prior work~\cite{Yu19,ElAtali24}.

\subsection{\BLACKOUT Software Stack}\label{sec:sw-stack}

We leverage Clang's existing \annotatetype attribute designed for static analysis tools~\cite{Braenne22} for the \blindedattribute attribute.
Normally, \annotatetype is not propagated to the LLVM \gls{ir}.
Thus, we extend Clang to emit \blindedattribute attributes into the \gls{ir} (\dCFour, \Cref{fig:swarch}).

We introduce an additional analysis pass for blinded variables (\dCFive, \Cref{fig:swarch}) in the LLVM-backend.
This pass performs recursive dependency analysis to trace all uses of \blindedattribute variables in data flows involving the \textsf{store}, \textsf{load}, and \gls{gep} \gls{ir} instructions. By recursively analyzing flows from blinded source instructions, e.g., loads from variables declared as \blindedattribute, the pass automatically propagates the blindedness property to variables acting as sinks, blinding them at their respective point of allocation.

We additionally introduce an instrumentation pass (\dCSix, \Cref{fig:swarch}) that transforms stack allocations (alloca \gls{ir} instructions) for variables annotated with \blindedattribute. The instrumentation adds the \texttt{llvm.cheri.cap.perms.and} intrinsic to each annotated alloca to unset the ``non-blinded'' permission control bit, thus turning the associated capabilitity into a \bc. Consequently, all stack memory associated with \blindedattribute variables are only accessible through \bcs.

\paragraph{\BC-enhanced allocator.}
We extended the Cornucopia revocation mechanism to provide a \blindedmalloc \gls{api} that returns \bcs to blinded heap allocations.
To meet \labelcref{req:overlap} (\Cref{sec:design}), such blinded heap allocations must not overlap with other, non-blinded allocations.
This ensures the \bc returned \blindedmalloc has exclusive access to the newly created allocation. The \blindedmalloc \gls{api} relies on Cornucopia to ensure the \bc does not overlap with any concurrently existing capabilities.
\gls{blackout}'s \blindedmalloc{} is integrated into CheriBSD via the \gls{mrs} (similar to the integration of Cornucopia, discussed in \cref{sec:cheri}).
This makes \bcs allocator-agnostic, allowing userspace processes to employ different underlying allocators as long as allocations occur via the shim.

\paragraph{Securely reclaiming blinded memory.}
Recall from \Cref{sec:challenges} that blinded memory must  be securely reclaimed to prevent information leaking from previously blinded memory regions.
\gls{blackout} implements explicit zeroing policies for blinded heap and stack allocations.
For blinded heap allocation, the \texttt{free} \gls{api}, implemented in the shim, inspects whether memory to be deallocated is blinded and erases the contents of blinded allocations before freeing it.
For stack-allocated blinded variables, the compiler automatically inserts a memset \gls{ir} instruction to zero out their memory immediately after the variable's lifetime ends.
This ensures sensitive data from blinded variables is securely erased, preventing unintended reuse or leakage when stack frames are reallocated.
Blinded global data persists until the process terminates and consequently does not need to be reclaimed.
The contents of (physical) pages are zeroed out by the \gls{os} before they are recycled for other processes.

\paragraph{Integration to CheriBSD.}
As discussed in \Cref{sec:hardware}, any newly created capability must initially be configured with its ``non-blinded'' permission bit set.
To support \bcs in CheriBSD, we perform a thorough scan of all CheriBSD code to ensure that any derived permissions (e.g., for drivers or userspace) also unset the ``non-blinded'' permissions. 
We also extend the CheriBSD \gls{cpu} exception and signal handlers to handle \bc exceptions and the new signal introduced to the \BLACKOUT \gls{cheri}-Toooba Core. 
We integrated \bcs into CheriBSD 24.05, resulting in a total of 100 lines of code changes over 14 distinct files.

\begin{table*}[t!]
    \centering
    \caption{Area and power costs on VCU118 @ 25MHz expressed in number of LUTs and registers, and Watts respectively.}\label{tab:hwcost}
    \vspace{-0.3cm}
    \begin{tabular}{r rcc rcc rcc rcc}\toprule
                                     & logic & \multicolumn{2}{c}{$\Delta (\%)$} &  memory & \multicolumn{2}{c}{$\Delta (\%)$} & registers & \multicolumn{2}{c}{$\Delta (\%)$} & power & \multicolumn{2}{c}{$\Delta (\%)$} \\ \midrule
        CHERI-Toooba Core            & 697508 & \multicolumn{2}{c}{--}      &   20852 & \multicolumn{2}{c}{--}       & 412493 & \multicolumn{2}{c}{--}      &   6.205 & \multicolumn{2}{c}{--} \\
        Blinded CHERI-Toooba Core    & 705863 & \multicolumn{2}{c}{1.2}     &   20855 & \multicolumn{2}{c}{0.0}      & 412913 & \multicolumn{2}{c}{0.1}     &   6.536 & \multicolumn{2}{c}{5.3} \\ \bottomrule
    \end{tabular}
\end{table*}

\section{Evaluation}\label{sec:eval}
We evaluate the overheads (area and performance) and security of \gls{blackout}. To evaluate overheads, we extended the CHERI-RISC-V Toooba FPGA softcore (RV64ACDFIMSUxCHERI), which is based on the open-source Bluespec RISC-V 64-bit Toooba core.

We use the BESSPIN-GFE security evaluation platform~\cite{Podhradsky22}, which has out-of-the-box support for the \gls{cheri}-Toooba softcore, replacing the standard core with our Blinded \gls{cheri}-Toooba. We synthesize the \gls{soc} at 25MHz (default for BESSPIN-GFE) targeting the Xilinx Virtex UltraScale+ VCU118 FPGA.

\subsection{Power \& resource usage}
\Cref{tab:hwcost} shows the power and resource usage obtained from Xilinx Vivado 2019.1. The overheads are minimal ($\approx$~1\% area and $\approx$~5\% power) compared to the unmodified CHERI-Toooba. This is expected since our hardware additions require no additional storage in memory or caches and only a single additional bit for registers.

\begin{table*}[t!]
    \centering
    \caption{Performance cost on VCU118 @ 25MHz expressed as CoreMark test results for $5x10^3$ iterations. The CoreMark score for a processor is reported as CoreMark-iterations-per-second-per-core-MHz. The $\Delta$ is relative to CHERI-Toooba Core results.}\label{tab:coremark}
    \vspace{-0.3cm}
    \begin{tabular}{r r rrc rrr rrr  c}\toprule
        \rowcolor{white}       &           &         Total ticks & \multicolumn{2}{c}{$\Delta$} &      Total time (sec) &   \multicolumn{2}{c}{$\Delta$} & Iterations/sec &  \multicolumn{2}{c}{$\Delta$} &         Score \\ \midrule
        \rowcolor{white} \multicolumn{2}{c}{\textbf{CHERI-Toooba}}&&&&&&&               \\
        \rowcolor{Gray!10}   & baseline (nocap) &         927674227 & \multicolumn{2}{c}{--} & 37 &         \multicolumn{2}{c}{--} &          135 &          \multicolumn{2}{c}{--} &           5.4 \\

        \rowcolor{white} & purecap  &       951983228 &          24309000 &          $2.62\%$ &          38 &           1 &          $3\%$&          131 &          4 &          $2.3\%$ &          5.24 \\ \midrule

        \rowcolor{white} \multicolumn{2}{c}{\textbf{Blinded CHERI-Toooba}}& & & & & & &&& \\
        \rowcolor{Gray!10}                & baseline (nocap) & 927729879 & 55652 &         $0.01\%$  &          37 &          0 &            $0\%$  &          135 &           0 &             $0\%$ &          5.4  \\
        \rowcolor{white}            & purecap       &          952083879 &   24409652
 &    $2.63\%$         &          38  &        1 &         $3\%$ &  131 &         4&       $2.3\%$ &    5.24 \\ \bottomrule  
\end{tabular}
\vspace{-0.1cm}
\end{table*}

\subsection{Performance}
For performance measurements, we run several benchmarks on the VCU118 FPGA.
First, we evaluate the impact of \BLACKOUT \gls{cheri}-Toooba on unblinded workloads by running the EEMBC CoreMark benchmark~\cite{gal12coremark} bare-metal on the GFE for $5\times10^3$ iterations over three separate test runs. \Cref{tab:coremark} shows the mean result demonstrating that there is only a negligible effect on performance for unblinded workloads between the unmodified \gls{cheri}-Toooba core and our \BLACKOUT \gls{cheri}-Toooba variant. 
We also obtain timing reports from Xilinx Vivado to show the effect of our hardware changes on the maximum clock frequency attainable. Both \gls{blackout} and the baseline are synthesizable at 25MHz, and the \gls{wns} for \gls{blackout} is $0.08$ns compared to $0.06$ns for the baseline, demonstrating no significant effect on clock frequency.

\paragraph{OISA benchmarks.} Second, for blinded workloads, we evaluate the performance impact of \BLACKOUT using five  OISA benchmarks adapted from Yu et al.~\cite{Yu19}. To adapt them to \BLACKOUT, we simply blind secret inputs with the \blindedattribute attribute, change dynamic allocations to use our blinded allocator \gls{api}, and compile the benchmarks using our \bc-enhanced compiler.
Porting the OISA benchmarks to \BLACKOUT took 1-5 LoC changes (<1\%) per benchmark.
We measured the performance on 
CheriBSD with \BLACKOUT support in pure-capability mode with the Cornucopia revocation mechanism enabled. Our setup follows the CheriBSD benchmark guide~\cite{DigitalSecuritybyDesign24}, with the exception of running the \BLACKOUT benchmarks in pure-capability mode (\Cref{sec:cheri}) with \bcs, and enabling the experimental Cornucopia revocation mechanism, which is essential to guarantee exclusive access for \bcs. We compiled all benchmarks with \texttt{-O3} and measured the combined user and system time using the \textit{time} command.
This measures the entire lifetime of blinded capabilities\textemdash including capability creation, algorithm execution, and memory reclamation (i.e., zeroing out blinded memory).
For the \texttt{findmax}, \texttt{binary\_search}, and \texttt{integer\_sort} benchmarks, we used blinded input arrays containing $2^N$ integers. For the matrix multiplication benchmark, we multiplied two square matrices of size
$2^{\frac{N}{2}} \times 2^{\frac{N}{2}}$. In the DNN example, the blinded input size was $2^{\frac{N}{2}}$, and the network consisted of two layers, each of size $2^{\frac{N}{2}} \times 2^{\frac{N}{2}}$, with a fixed output size of $2^6$. We evaluated all benchmarks with
$N \in \{12, 14, 16, 18, 20\}.$

The results in \Cref{fig:benchmarks} show the OISA benchmark run-time in nanoseconds for different input sizes and three configurations: \textit{baseline}, with capability-enforcement disabled, \textit{purecap}, which enforces CHERI memory-safety only, and \textit{purecap + blinded}, which enforces all \gls{blackout} invariants including CHERI memory-safety and data-obliviousness. As our hardware modifications do not add additional cycles to any instructions, all configurations can run with the same FPGA bitstream, requiring only different compilation flags. The baseline configuration uses \gls{cheri}'s ``hybrid'' mode which allows \gls{cheri}-capable processors to run legacy, non-capability code.

The geometric mean overhead per benchmark is shown in \Cref{fig:overheadtable}. The overall result shows a minimal geometric mean overhead of \OverheadRelativeToCheri{} for \gls{blackout} compared to the \textit{purecap} configuration, which is several times lower than overheads for prior work enforcing data-oblivious computation~\cite{Yu19,ElAtali24}.
The overhead in blinded workloads is caused by several factors:
\begin{enumerate}[topsep=0pt, leftmargin=11pt]
    \item Clearing blinded memory at the end of a blinded capability's lifecycle. This is done by the compiler for blinded memory on the stack on function returns, and by the \blindedmalloc for blinded memory on the heap on freeing. For instance, when N = 20, binary search operates on a 4MiB array. In this setting, memory reclamation (zeroing) accounts for most of the overhead.
    \item Additional stores and loads to set the blindedness bit in registers. When initializing blinded data at the start of the benchmarks, any data in registers must be first stored into blinded memory and then loaded back to set the register blindedness bit. This only occurs for the initial blinded data and is not required to propagate blindedness during execution. The effect on performance is therefore minimal. Nevertheless, we discuss a potential optimization to avoid this in \Cref{sec:blindedInst}.
    \item Changes to instruction cache performance caused by the slight increase in code size due to compiler instrumentation.
\end{enumerate}

\definecolor{baselinecolor}{HTML}{1f77b4}
\definecolor{purecapcolor}{HTML}{ff7f0e}
\definecolor{purecapblindedcolor}{HTML}{2ca02c}
\newcommand\baseblock{\colorbox{baselinecolor}{  }}
\newcommand\pcblock{\colorbox{purecapcolor}{  }}
\newcommand\pcblndblock{\colorbox{purecapblindedcolor}{  }}
\begin{figure*}[t]
\footnotesize
\begin{centering}
    \begin{subfigure}{0.28\linewidth}
    \centering
        \noindent\includegraphics[height=4cm]{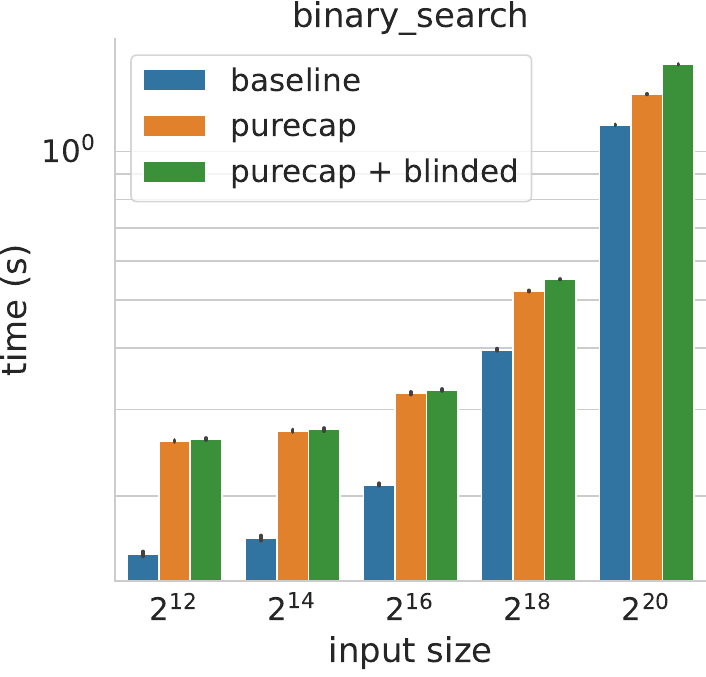}
        \caption{}
    \end{subfigure}
    \begin{subfigure}{0.28\linewidth}
    \centering
        \noindent\includegraphics[height=4cm]{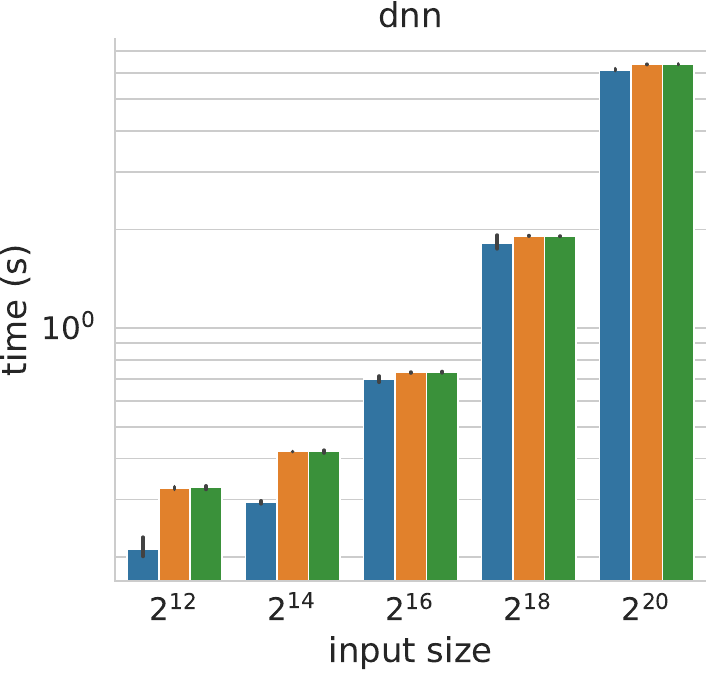}
        \caption{}
    \end{subfigure}
    \begin{subfigure}{0.28\linewidth}
    \centering
        \noindent\includegraphics[height=4cm]{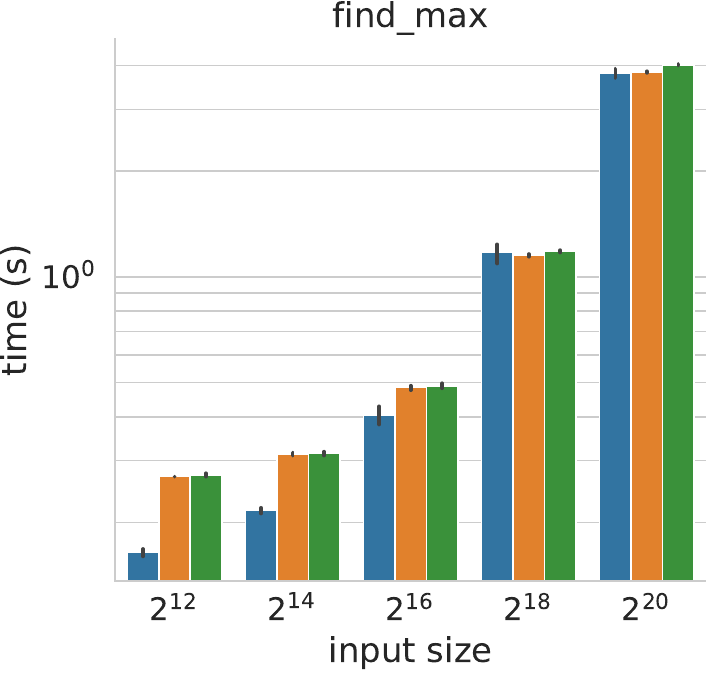}
        \caption{}
    \end{subfigure}
    \begin{subfigure}{0.28\linewidth}
    \centering
        \noindent\includegraphics[height=4cm]{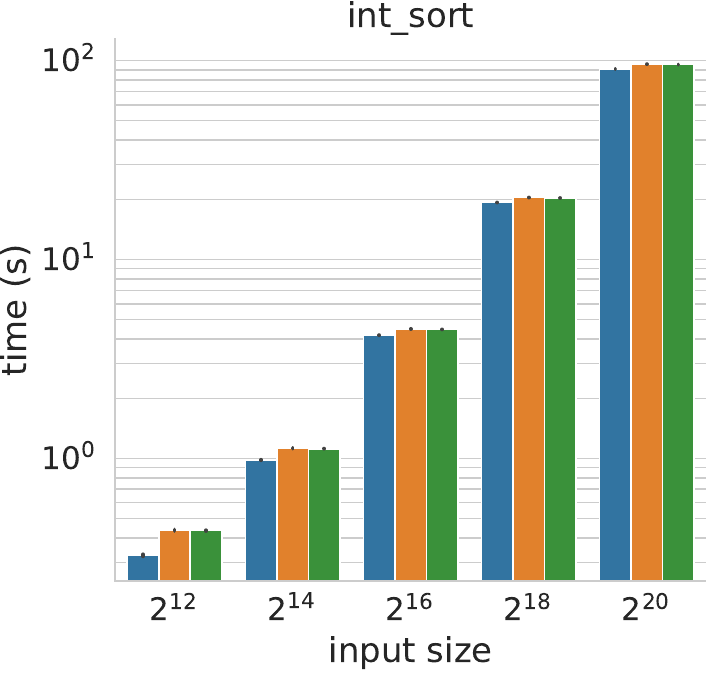}
        \caption{}
    \end{subfigure}
    \begin{subfigure}{0.28\linewidth}
    \centering
        \noindent\includegraphics[height=4cm]{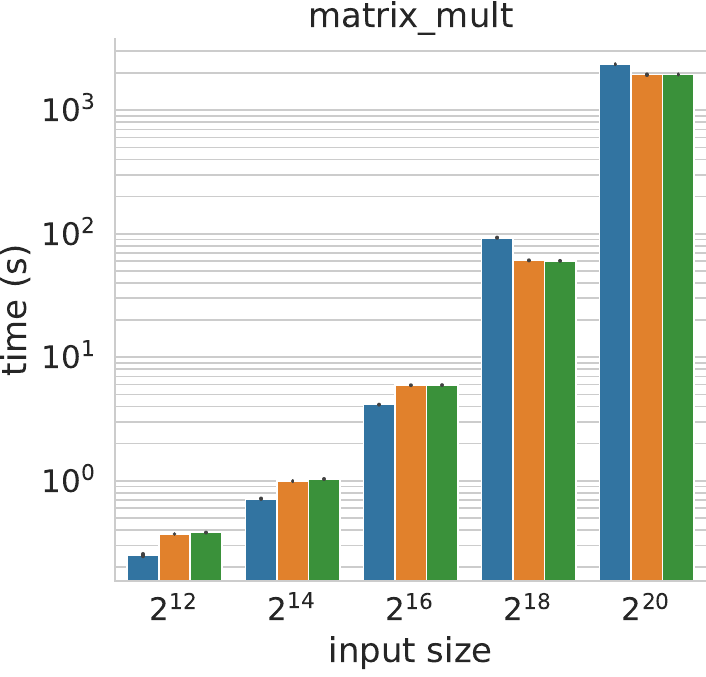}
        \caption{}\label{fig:overheadmatrixmult}
    \end{subfigure}
    \begin{subfigure}{0.28\linewidth}
    \centering
        \raisebox{2cm}{
        \begin{tabular}{r rrr}\toprule
                                       & \multicolumn{3}{c}{Overhead (\%)} \\ \cmidrule(lr){2-4}
            \multirow{2}{*}{Benchmark} & \multirow{2}{*}{\(\frac{\raisebox{0.5em}{\pcblndblock{}}}{\pcblock{}}\)}
                                       & \multirow{2}{*}{\(\frac{\raisebox{0.5em}{\pcblock{}}}{\baseblock{}}\)}
                                       & \multirow{2}{*}{\(\frac{\raisebox{0.5em}{\pcblndblock{}}}{\baseblock{}}\)} \\ 
                            &       &       &                       \\ \midrule
            binary\_search  & 4.5   & 45.5  & 52.1                  \\
            dnn             & 0.0   & 20.1  & 20.2                  \\
            find\_max       & 2.0   & 23.0  & 25.5                  \\
            int\_sort       & -0.5  & 13.1  & 12.5                  \\
            matrix\_mult    & 1.3   & 9.7   & 11.1                  \\ \bottomrule
        \end{tabular}
        }
        \caption{}\label{fig:overheadtable}
    \end{subfigure}
    \vspace{-0.4cm}
    \caption{Run-time in seconds of the OISA \texttt{binary\_search} (a), \texttt{dnn} (b),\texttt{find\_max} (c), \texttt{int\_sort} (d), and \texttt{matrix\_mult} (e) benchmarks built for the \emph{baseline} with capability-enforcement disabled, \emph{purecap} mode with \gls{cheri} memory-safety enforced but data-obliviousness not enforced, and \emph{purecap+blinded} mode with both \gls{cheri} memory-safety and data-obliviousness enforced. \Cref{fig:overheadtable} shows the geometric mean overheads for each benchmark aggregated over all input sizes.}\label{fig:benchmarks}
\end{centering}
\end{figure*}

\begin{figure*}[t]
\begin{centering}
    \begin{subfigure}{0.28\linewidth}
    \centering
        \noindent\includegraphics[height=4cm]{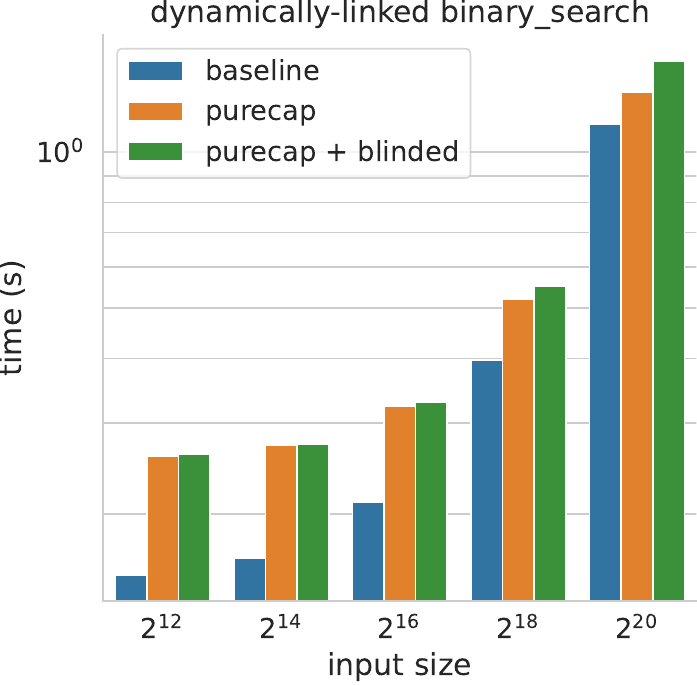}
        \caption{}
    \end{subfigure}
    \begin{subfigure}{0.28\linewidth}
    \centering
        \noindent\includegraphics[height=4cm]{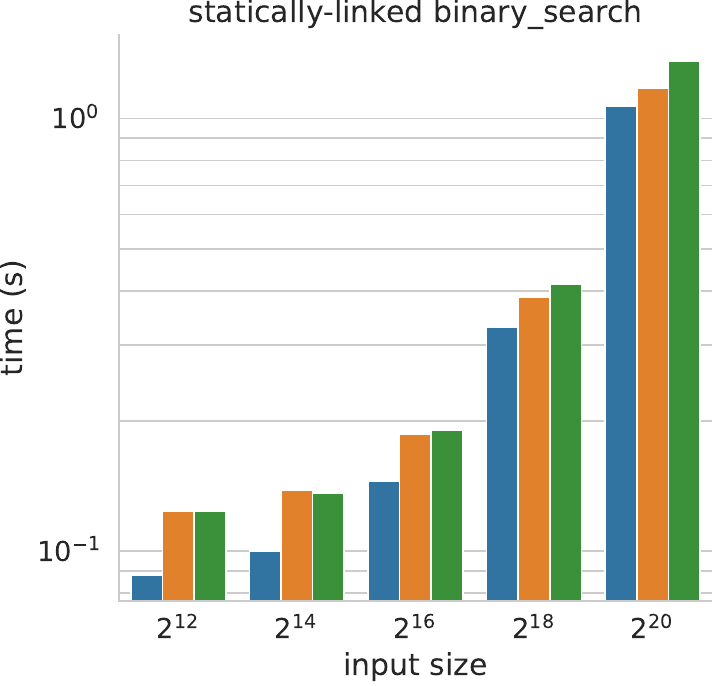}
        \caption{}
    \end{subfigure}
    \begin{subfigure}{0.28\linewidth}
    \centering
        \noindent\includegraphics[height=4cm]{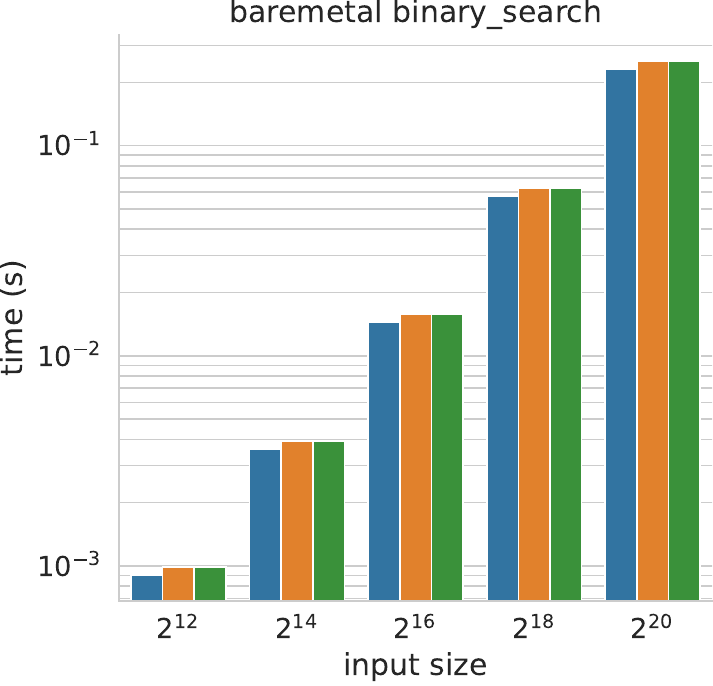}
        \caption{}
    \end{subfigure}
    \vspace{-0.4cm}
    \caption{Run-time in seconds of the OISA \texttt{binary\_search} benchmark in dynamically-linked, statically-linked, and bare-metal configurations for the \emph{baseline}, \emph{purecap}, and \emph{purecap+blinded} modes. The geometric mean overhead of \emph{purecap+blinded} compared to \emph{baseline} aggregated over all input sizes falls from $52.1\%$ for the dynamically-linked configuration (a) to $32.0\%$ for the statically-linked one (b). The corresponding geometric mean is $9.1\%$ for the baremetal configuration (c).}\label{fig:binsearchbenchmarks}
\end{centering}
\vspace{-1em}
\end{figure*}

We also evaluate the \emph{combined} performance impact of \gls{cheri}'s memory-safety and \BLACKOUT's data-oblivious enforcement.
We measure a moderate geometric mean overhead of \OverheadRelativeTobaseline compared to the unprotected baseline configuration with capability-enforcement completely disabled.
We observe that the overhead for both \gls{cheri}'s purecap and \BLACKOUT's purecap + blinded modes relative to the unprotected baseline is significantly larger in the \texttt{binary\_search} benchmark, particularly for smaller input sizes, compared to the other OISA benchmarks. We investigate further the composition of the overhead in that benchmark.
In consultation with \xblackout{the University of Cambridge Computer Laboratory Security Group} we identified three possible reasons for the poor performance:
\begin{enumerate}[topsep=0pt, leftmargin=10pt]
  \item Inefficiencies in CHERI-RISC-V’s relocation representation and processing converting function pointers into capabilities.\label{lim:reloc}
  \item Lack of support for lazy binding of functions in CHERI-RISC-V which manifests as high initial load times at startup.\label{lim:no-lazy-binding} 
  \item The current version of CHERI-LLVM currently disables most loop optimizations, making it slower for the type of code the \texttt{binary\_search} benchmark relies on compared to compilation to a non-\gls{cheri} RISC-V target.\label{lim:no-loop-opt} 
\end{enumerate}

To isolate the effect of these potential root causes for the overhead that are independent of the \BLACKOUT-related changes, we repeat the \texttt{binary\_search} benchmark in two additional configurations:
\begin{inparaenum}[a)]
  \item a statically linked benchmark on CheriBSD designed to mitigate the inefficiencies in relocation (\labelcref{lim:reloc}) and lazy binding (\labelcref{lim:no-lazy-binding}), and
  \item a bare-metal version of the benchmark that runs without CheriBSD. 
\end{inparaenum}
Crucially, the bare-metal benchmark measurements include only the execution of the data-oblivious algorithm, eliminating the initial startup cost, reclaiming of memory, and system calls, thus isolating the computational cost of \bcs without their impact on memory management.

\Cref{fig:binsearchbenchmarks} shows the results of these additional \texttt{binary\_search} configurations and allows us to make the following observations:
First, in the statically-linked benchmarks, the overhead for smaller input sizes improve, but the result for larger input sizes is similar to the dynamically-linked benchmarks, suggesting that the one-off cost of initial load times is amortized over the longer benchmark runs in both cases.
Second, the overheads for both purecap and purecap + blinded modes compared to the baseline are significantly lower when run on baremetal. This supports our hypothesis that relocation (\labelcref{lim:reloc}) and lack of lazy binding (\labelcref{lim:no-lazy-binding}) are the main sources of overhead for \texttt{binary\_search} in \Cref{fig:benchmarks}.

We also note some cases in the \texttt{matrix\_mult} benchmark in \Cref{fig:overheadmatrixmult} where the baseline performs slightly worse than the other two. We examined the generated assembly for these benchmarks and discovered that the \gls{cheri}-enabled compiler does a better job at optimizing parts of the code whose effect on performance grows with input size, leading to slightly lower run-times in the purecap and purecap + blinded configurations.

\paragraph{SpectreGuard benchmarks.} Third, we run the SpectreGuard~\cite{Fustos19} synthetic benchmark using the version adapted by Daniel et al.~\cite{Daniel23} to evaluate ProSpeCT (see \Cref{sec:relatedwork}).
This version uses data-oblivious implementations of the \texttt{chacha20}, \texttt{sha2}, and \texttt{curve25519} cryptographic algorithms from HACL*\cite{Zinzindohoue17}, a formally verified cryptographic library written in F*~\cite{Swamy16}.
The benchmarks represent a workload consisting of public-data computations, which gain substantial performance from speculative execution, alongside encryption routines, which are less impacted by speculation. Each benchmark varies the proportion of speculative versus cryptographic work (S/C), as shown in \Cref{tab:spectreguard}.
The ProSpect benchmark version already annotates all secret values in those test cases.
We adapt those annotations to \bas and run all the test cases on \BLACKOUT CHERI-Toooba in both \emph{purecap} and \emph{purecap + blinded} versions.
\Cref{tab:spectreguard} shows the results (average of 20 runs) for the \texttt{chacha20} and \texttt{sha2} test cases.
Unlike for the OISA benchmark results, where \BLACKOUT's overhead could be attributed to zeroing out stack and heap variables, the ProSpeCT version of the SpectreGuard benchmark places all secret variables in static variables allocated from the program's data section, similar to global variables.
Consequently, \BLACKOUT does not introduce any discernible overhead for encryption time or workload time.
The slight speed improvement of the purecap +  blinded version compared to the purecap baseline in chacha20 25S/7C and sha2 90S/10C falls within standard deviation (maximum $\sigma = 0.08\%$).
For completeness, we also include results for nocap, which performs worse than purecap in most cases. We attribute this difference to improvements in the RISC-V backend for \gls{cheri}-RISC-V targets compared to plain RISC-V targets in \gls{cheri}-LLVM, resulting in more compact and efficient assembly. We confirmed this through manual investigation.

For the Curve25519 test case, we discovered a constant-time violation caused by a secret-dependent branch instruction in the compiled binary.
Although this violation is not present in the formally verified source code, it is introduced by the compiler, which in our case is CHERI’s fork of Clang.
This violation is not reported in ProSpeCT’s evaluation because it uses the GCC compiler, which likely correctly avoided generating such a secret-dependent branch.
This difference in generated assembly code highlights the usefulness of \BLACKOUT: different compilers and/or configurations can silently introduce constant-time violations which \BLACKOUT can detect and stop even if the source does not contain such a violation.

\begin{table*}[h]
\centering
\caption{SpectreGuard benchmark performance average measured over 20 runs of each test case (maximum $\sigma = 0.08\%$).}
\vspace{-0.3cm}
\resizebox{\textwidth}{!}{
\begin{tabular}{lcccccccc}
\toprule
 \multirow{3}{*}{\textbf{Blinded CHERI-Toooba}} & \multicolumn{2}{c}{\textbf{25S/75C}}  &  \multicolumn{2}{c}{\textbf{50S/50C}} & \multicolumn{2}{c}{\textbf{75S/25C}}  & \multicolumn{2}{c}{\textbf{90S/10C}} \\
 \cmidrule(lr){2-3}\cmidrule(lr){4-5}\cmidrule(lr){6-7}\cmidrule(lr){8-9}
&   Total Ticks & $\Delta$ (to row above) & Total Ticks & $\Delta$ (to row above) &  Total Ticks & $\Delta$ (to row above) & Total Ticks &  $\Delta$ (to row above) \\
\midrule
\textbf{chacha20} & & & & &&&& \\
nocap  & 25860874 & -- & 23132682 &   -- & 25417083 & -- & 20661078 & -- \\
\rowcolor{Gray!10} purecap  & 22546871 & -3314003 ($-12.81\%$) & 20378637  & -2754045 ($-11.91\%$) & 25430015 & 12932 ($0.05\%$) & 20256607 & -404472 ($-1.96\%$) \\
purecap + blinded  & 22541024 & -5847 ($-0.03\%$)& 20378935 &   298 (0.00\%) & 25434929 & 4914 ($0.02\%$)& 20256061 & -546 ($0.00\%$) \\
\bottomrule
\textbf{sha2} & & & & &&&& \\
nocap  & 24819790 & -- & 21595015 & -- & 25417599 & -- & 22163550 & -- \\
\rowcolor{Gray!10} purecap  & 24771185 & -48604 ($-0.20\%$) & 21591164 & -3851 ($-0.02\%$) & 26342237 & 924638 ($3.64\%$) & 22378912 & 215362 ($0.97\%$) \\
purecap + blinded  & 24771730 &  544 ($0.00\%$) & 21593499 & 2335 ($0.01\%$) &26342910 & 672 ($0.00\%$)& 22371464 & -7448 ($-0.03\%$)\\
\bottomrule
\end{tabular}
}
\label{tab:spectreguard}
\vspace{-1em}
\end{table*}

\enlargethispage{\baselineskip}
\subsection{Security}\label{sec:sec-eval}
\paragraph{Empirical Spectre mitigation evaluation.}
CHERI-Toooba is vulnerable to Spectre \gls{btb}~\cite{Kocher19}, \gls{rsb}~\cite{Maisuradze18,Koruyeh18} and \gls{stl}~\cite{Horn18} (cf. test suite in \cite{Fuchs21a}). We blinded the secret value in all test cases by inserting a single \texttt{candperm} instruction. As seen in \Cref{tab:transientattack}, \BLACKOUT prevents all Spectre attacks from \cite{Fuchs21a} because \BLACKOUT catches side-channel violations even during speculation, but suppresses faults until speculation is confirmed (and ignores them otherwise). Additionally, we have replicated ProSpeCT's~\cite{Daniel23} Spectre tests.
However, as they involve dereferencing a secret data value as a pointer, they are inherently prevented by CHERI's architectural security properties, which preclude dereferencing non-capability data. 
Note that none of the Spectre tests in \cite{Fuchs21a,Daniel23} use transient capability forgery, and therefore do not require \gls{capSpecCon} to prevent.

\begin{table}
\caption{Transient-execution attacks successfully prevented
on CHERI-Toooba and Blinded CHERI-Toooba. \checkmark{} indicates the attack is successfully defended against, while \xmark{} indicates the attack succeeds.}
\vspace{-0.3cm}
\label{tab:transientattack}
\centering
\resizebox{0.85\columnwidth}{!}{
\footnotesize
\begin{tabular}{lcccc}
\toprule
 & \multicolumn{4}{c}{Spectre variant} \\
\cmidrule(lr){2-5}
 & PHT & BTB & RSB & STL \\
\midrule
\textbf{CHERI-Toooba} & \checkmark & \xmark & \xmark & \xmark \\
\textbf{Blinded CHERI-Toooba} & \checkmark & \checkmark & \checkmark & \checkmark \\
\bottomrule
\end{tabular}}
\vspace{-1.2em}
\end{table}

\paragraph{Empirical non-interference evaluation.} Data-oblivious software inherently provides the non-interference property~\cite{Winderix24}.
We verify this empirically using the methodology from \cite{Winderix24, Winderix24a, Bognar23}: relevant signal traces are extracted from the \gls{vcd} waveforms across different program runs to ensure that runs with different secret data values produce identical signal traces.
As in prior work~\cite{Winderix24}, we extract traces for signals that could leak secret data through cache-timing (addresses of cache accesses), speculative execution (branch predictor states), and port contention and instruction latencies (reorder buffer scheduling).
As \BLACKOUT enforces data-oblivious processing of blinded data, we use the data-oblivious \texttt{binary\_search} program from \cite{Yu19} as a case study. We vary the blinded data values between two runs of the program, and verify that the extracted traces for the relevant signals are identical.

\paragraph{Theoretical security evaluation.} Our security argument is composed of three invariants:
\begin{enumerate}[topsep=0pt, leftmargin=10pt]
    \item \textbf{Standard CHERI-provided memory safety}. This includes attacks based on memory safety violations, such as out-of-bounds access and use-after-free. We do not loosen any of the restrictions imposed by standard CHERI (such as proper capability bounds and lifetime). We are therefore compliant with the standard CHERI model, and as such inherit its safety guarantees against spatial and temporal memory violations. This extends to the guarantees provided by formal models~\cite{Gao21,DuqueAnton25,Ploix25}.
    \item \textbf{Side-channel protection for blinded data}. This is provided by the enforcement of data-oblivious computation on blinded data (\labelcref{req:loadstore,req:controlflow}). Any violation of this results in a fault, as explained in \Cref{sec:design}. We inherit formal guarantees for this from the OISA and BliMe models~\cite{Yu19,ElAtali24}. We further show its efficacy empirically by running the OISA benchmarks~\cite{Yu19} using \BLACKOUT and verify that introducing non-data-oblivious changes to them is either detected by our compiler or causes a run-time fault.
    \item \textbf{Confidentiality of blinded data with respect to direct access}. While BliMe guarantees the confidentiality of blinded data in memory using in-memory tags, we achieve the same goal by ensuring that 1) blinded capabilities have exclusive access to blinded data throughout their lifetime (\labelcref{req:overlap}), 2) memory containing blinded data is cleared at the end of the corresponding blinded capability's lifetime (also \labelcref{req:overlap}), and 3) blinded data cannot be stored to memory using unblinded capabilities (\labelcref{req:blindedstore}). The resulting combination ensures that, as in OISA and BliMe, any data stored as blinded into memory, is loaded as blinded into registers. This holds until the data is explicitly declassified (\Cref{sec:design}).
\end{enumerate}
 \enlargethispage{\baselineskip}

\section{Discussion \& Limitations}\label{sec:discussion}

\paragraph{Blinding existing variables and oblivious-data-race-safety.}
As discussed in \Cref{sec:design}, \BLACKOUT ensures that memory designated as blinded---either annotated by the developer, or inferred as such by the compiler---is blinded at allocation time.
This ensures exclusive access through corresponding \bcs, satisfying invariants \labelcref{req:blindedstore,req:noblindedcap,req:overlap} (\Cref{sec:design}).
However, some correct data-oblivious programs may initially allocate memory as non-blinded, only requiring it to be blinded after interacting with other blinded data.
While \BLACKOUT restricts this pattern, a less-restrictive variation could allow dynamically blinding previously allocated memory. 
This would require scanning program memory to revoke existing, non-blinded capabilities referencing to-be-blinded memory, similar to revocation strategies used by Cornucopia~\cite{WesleyFilardo20} and Cornucopia Reloaded~\cite{Filardo24}.
The trade-off is increased performance overhead. 
Future work might explore dynamic or delayed revocation strategies to reduce this cost while ensuring eventual exclusive access.

Eventually, data-oblivious software must \emph{unblind} memory after data-oblivious computation concludes and secret intermediates are cleared (\Cref{sec:design}).
In concurrent scenarios, shared blinded memory may require revoking overlapping \bcs to maintain temporal memory safety and prevent unauthorized reuse after unblinding.
Currently, \BLACKOUT does not address this \emph{oblivious-data-race-safety}, leaving it as an open problem for future work.

\paragraph{Unblinding results through capability escrow.}\label{sec:capescrow}
A solution for further securing the unblinding of \bd containing non-sensitive results is to leverage \gls{cheri}'s facilities for \emph{capability sealing} (\Cref{sec:cheri}).
Sealed capabilities cannot be de-referenced and protect against tampering by fixing properties like permissions and bounds; any attempt to tamper with the capability will result in invalidating it.
In this context, sealed capabilities enable a form of ``\emph{capability escrow}'', where a non-blinded version of a \bc stored for safe-keeping until the data-oblivious operations complete.
Software can use a \bc to seal its non-blinded counterpart, which cannot be de-referenced while sealed and thus poses no threat to confidentiality.
After data-oblivious computation has completed, secret intermediate values are cleared, and the \bc is used to unseal the non-blinded counterpart, allowing access to the result.
Designing software \glspl{api} for capability escrow is left for future work.

\paragraph{Different techniques for developer annotations}
In \BLACKOUT, developers use \blindedattribute annotations to indicate blinded variables (\Cref{sec:sw-stack}).
Compiler attributes are a common mechanism for conveying semantic information without changing the underlying language.
\BLACKOUT leverages Clang's existing \annotatetype mechanism, available since LLVM 14, avoiding intrusive compiler changes.

However, this approach has limitations---annotating positional parameters in functions can be cumbersome within the confines of existing types of annotations.
An alternative would be to introduce a ``blinded C'' dialect that integrates blinding directly into the type system, similar to the approaches discussed for Rust in \Cref{sec:intro}. 
For example, a \texttt{blinded} keyword akin to \texttt{const} could be used to mark variables as a blinded counterpart of their basic type.
While this would improve developer ergonomics and simplify function parameter annotations, it would require intrusive changes to the language and compiler. This could hinder maintainability and extensibility, especially for evolving platforms like CHERI.

\paragraph{Limitations of \gls{brr} structures for blinded data storage.}
\glsreset{brr}
While \gls{brr} structures effectively manage blinded register spills by clearly identifying and restoring blinded values, it is impractical to store all blinded data exclusively using \gls{brr}-like structures.
The principal reason is the substantial memory overhead\textemdash \gls{brr} structures effectively double memory consumption.
Such overhead would significantly degrade system performance and scalability, particularly for applications requiring large quantities of blinded data. Therefore, \gls{brr} structures should remain reserved for specific contexts, such as register spilling, while more efficient \bcs handle general \bd storage.

\paragraph{\gls{isa} extensions.}\label{sec:blindedInst}
As we note in \Cref{sec:hardware}, \BLACKOUT, unlike previous data-oblivious \glspl{isa}~\cite{Yu19,ElAtali24}, does not add any additional instructions to the underlying \gls{isa} but leverages existing \gls{cheri} functionality with only small adjustments to its permission model to facilitate necessary architectural changes to support \bcs.
Future work can extend the \gls{isa} by adding dedicated instructions to support \bd.
A potential extension is an instruction to directly set the blindedness bit in a register without a load via a \bc.
This would allow the \bc-enhanced compiler to optimize code where a certain variable can be kept completely in a register throughout its lifetime.
Currently, \BLACKOUT requires such variables to be allocated on the stack in order to load them via the corresponding \bc.

\paragraph{Relevance for \gls{cheri} standarization.}
\BLACKOUT demonstrates a practical method for enhancing \gls{cheri}'s security model without invasive architectural changes by integrating data-oblivious computation capabilities into the existing \gls{cheri} protection model.
This integration provides insights into how capability-based architectures can evolve to support advanced security properties, such as data-oblivious computation, while maintaining compatibility and performance.
Consequently, \BLACKOUT's principles and mechanisms can inform
ongoing \gls{cheri} standardization efforts~\cite{CheriAlliance25}, potentially guiding the evolution of \gls{cheri}-enabled architectures towards broader security guarantees. Above, we discussed adding custom instructions. But maintaining the \gls{cheri} \gls{isa} as we currently do in \BLACKOUT provides backward-compatibility for non-\BLACKOUT hardware and can ease integration into the standard.

\paragraph{Relationship to non-interference.}
Since \BLACKOUT builds on BliMe, it also ensures (as shown in Section VI of \cite{ElAtali24}) the values of blinded data have no effect on the rest of the system by 
\begin{inparaenum}[1)]
    \item preventing blinded values from directly flowing to insecure instructions, and 
    \item by ensuring no secret-dependent branches are allowed when operating on blinded data, with an exception for controlled declassification as described above and in \Cref{sec:design-hw-arch}. 
\end{inparaenum}
Given its scope, we conjecture that this guarantee satisfies the non-interference property even though it was not explicitly discussed in \cite{ElAtali24}. Providing a formal model for \BLACKOUT itself and explicitly establishing non-interference are left for future work. Nevertheless, we empirically investigate non-interference for \BLACKOUT in \Cref{sec:sec-eval}.

\section{Related Work}\label{sec:relatedwork}

\paragraph{Developing side-channel resistant software.}
Software that handles sensitive data---such as cryptographic libraries or \glshyph{os} kernels---must carefully avoid introducing side-channel leakage.
Most code is not naturally resistant to such leaks as writing constant-time code is often counterintuitive for programmers used to optimizing for performance or resource efficiency~\cite{Intel19}.
Development toolchains are not well suited for constant-time programming for two main reasons:
\begin{inparaenum}
\item optimizing compilers can break data-obliviousness properties of high-level code~\cite{Schneider24}, and
\item constant-time libraries~\cite{GoAuthors25,lovecruft25,McLean23a,klutzy15} must balance security with the cost of disabling optimizations, e.g., by implementing constant-time functionality in architecture-specific inline assembly.
\end{inparaenum}
\ifnotabridged
A complete solution would likely require invasive changes to the compiler infrastructure, such as extending LLVM’s type system to annotate values needing timing protection and limiting optimizations on them---at significant performance cost~\cite{McLean23a}.
Some constant-time Rust libraries avoid compiler optimization interference by implementing all sensitive operations in inline assembly~\cite{klutzy15,McLean23}, which is opaque to LLVM's optimization passes.
However, this also prevents the Rust compiler from verifying the code's memory-safety correctness and inhibits even simple optimizations such as constant folding and algebraic simplification.
\fi

\paragraph{Verifying constant-time code.} Researchers have explored various methods for writing and verifying data-oblivious and constant-time code~\cite{Molnar05,Bernstein05,Bernstein06,Coppens09,Cleemput12,Andrysco15,Rane15,Ohrimenko16,Fisch17,Shaon17,Zheng17,Ahmad18,Mishra18,Sasy18,Eskandarian19,Borrello21,ElAtali24a}, resulting in numerous tools offering informal~\cite{Reparaz17,Weiser18,Wichelmann18,Wichelmann22,Disselkoen24,Langley25} and formal~\cite{Cauligi17,Daniel20} guarantees.
An actively maintained list of ``constant-timeness'' verification tools (CT-tools) currently includes 55 such tools~\cite{MasarykUniversityCentreforResearchonCryptographyandSecurity25}.
However, both static analysis or formal methods face signficant challenges.
The principal shortcoming of static analysis approaches is that data-obliviousness can only be defined at machine-code level, rather than for high-level language constructs.
Capturing microarchitectural subtleties of real-world hardware in formal models (and keeping the models up-to-date as hardware evolves) is difficult.
Static analyses may not be sound and can lead to over- or under-tainting.
Lastly, testing whether a program is data-oblivious remains challenging as tools built on static and formal analysis are typically not integrated into modern toolchains, have significant technical limitations including high overheads to compilation time, many false-positives, and are difficult to use~\cite{Jancar22,Geimer23,Fourne24}.
 
In practice, constant-time coding practices alone are unreliable~\cite{Pornin25}.
Compiler optimizations continue to evolve, and new compilers  emerge in unexpected contexts (e.g., in-silicon \glsdesc{jit} compilers in \gls{cpu} hardware).
Documentation gaps across the software and hardware stack further hinder efforts, especially when target platforms are not narrowly confined~\cite{Pornin18}. 
Thus, robust enforcement of data-obliviousness requires hardware/software codesign~\cite{Liu15,Yu19,ElAtali24,ElAtali24a}.
Such designs allow programmers to annotate sensitive data, enabling hardware to enforce protection against side-channel leakage.
Data-oblivious \glspl{isa}~\cite{Yu19,ElAtali24}, discussed in \Cref{bg:data-oblivious-isa}, use hardware-based taint tracking to transform \emph{silent} timing side-channel leaks into \emph{explicit} hardware faults.
Yet, programming for data-oblivious \glspl{isa} remains hard due to the lack of integrated toolchain support for expressing such software/hardware contracts.

\Gls{ami}~\cite{Winderix24} introduces new \gls{isa} primitives for more efficient control-flow-linearization, addressing how to make programs data-oblivious.
\gls{ami} is complementary to data-oblivious \glspl{isa}, which enforce data-obliviousness through \gls{isa} contracts.
However, it requires greater developer involvement as it requires developer write assembly manually (with correct usage of new primitives); incorrect usage can \emph{silently} leak secrets.

\BLACKOUT complements static-analysis and formal verification approaches by bridging the gap between high-level verified data-oblivious software and the low-level machine code.
\BLACKOUT \emph{enforces} data-oblivious implementation through a hardware/software co-design that integrates with mainstream toolchains.
Unlike static analysis tools, it provides hardware-enforced taint tracking, reliably converting implicit timing  leaks into explicit faults.
By extending the \gls{cheri} capability model with \bcs, \BLACKOUT ensures both memory-safety and data-obliviousness without invasive architectural changes.
It simplifies secure programming by modifying the CHERI Clang/LLVM compiler to infer when data-oblivious invariants should be enforced.
Thus, \BLACKOUT simultaneously addresses usability, security, and performance challenges that previous approaches tackled in isolation.

\paragraph{Securing speculation.}
Many hardware defenses adapt \gls{cpu} microarchitectures to defend against Spectre and other speculative execution side-channel attacks by isolating the microarchitectural elements, such as cache hierarchies~~\cite{Kiriansky18,Yan18,Sakalis19,Saileshwar19,Taram19,Ainsworth20,Kim20,Zhao20,Loughlin21}, that can either influence speculation or leak data across security domains.
Speculative taint-tracking techniques\cite{Yu19a, Fustos19, Schwarz20, Daniel23} delay instructions dependent on speculatively loaded secret data. 
Of recent work, ProSpeCT~\cite{Daniel23} formalizes the constant-time policy with respect to control flow and memory accesses for a broad class of speculation mechanisms.
Its scope is guaranteeing that hardware does not leak secrets of constant-time programs during speculation.
Unlike data-oblivious \glspl{isa}, neither ProSpeCT nor other transient execution defenses enforce their guarantees on non-speculative execution.
As such, they do not help developers write constant-time code.
\BLACKOUT prevents secret leakage even when programs are not constant-time and helps developers refine annotations.
Future work can integrate formal approaches with \BLACKOUT so that once developers transform code, it can be guaranteed data-oblivious.

\section{Conclusion}

We introduced \BLACKOUT, a novel approach that integrates data-oblivious computation into the \gls{cheri} capability-based security model.
Future research will focus on optimizing performance of reclaiming \bd, exploring dynamic revocation techniques, and bringing these into \gls{cheri} standardization efforts.

\section*{Acknowledgments}
We thank our colleagues at \xblackout{University of Waterloo} and \xblackout{Ericsson}:
\xblackout{Adam Caulfield}, \xblackout{Christoph Baumann},
\xblackout{H\aa kan Englund}, \xblackout{Santeri} \xblackout{Paa-\\volainen}, and
\xblackout{Sini Ruohomaa} for their comments on drafts of this \ifanonymous manuscript\else paper\fi.
We also thank the 
\xblackout{University of Cambridge Computer Laboratory} members, particularly \xblackout{Jessica Clarke},
\xblackout{Jonathan Woodruff}, \xblackout{Peter Rugg}, \xblackout{Robert Watson}, and former member \xblackout{Alex Richardson}.

This work is supported in part by \xblackout{Natural Sciences and Engineering Research Council of Canada} (grant number \xblackout{RGPIN-2020-04744}), and the \xblackout{Government of Ontario (RE011-038)}. Views expressed in the paper are those of
the authors and do not necessarily reflect the position of \xblackout{the
funders}.

\bibliographystyle{ACM-Reference-Format}
\bibliography{main.bib,local.bib}
\end{document}